\documentclass[fleqn]{article}
\date{}
\usepackage[dvips]{graphicx}  
\begin{document}
.
\begin{center}

\vspace{20mm}

{\LARGE \bf THE GLOBAL COLOUR MODEL OF QCD FOR HADRONIC PROCESSES - A REVIEW
\footnote{Published in Fizika B {\bf 7}(1998)171.}}

\vspace{3cm}
{{\Large Reginald T. Cahill  and      Susan M. Gunner
  \footnote{E-mail: Reg.Cahill@flinders.edu.au,
Susan.Gunner@flinders.edu.au}}

\vspace{1cm}
  {\large Department of Physics, Flinders University}

 {\large GPO Box 2100, Adelaide 5001,
Australia }}

\vspace{1cm}

\begin{minipage}{120mm}
\begin{center}{\bf Abstract}\end{center}
{The Global Colour Model (GCM) of QCD is a quark-gluon quantum field theory
that very successfully  models QCD for  low energy  hadronic processes. 
An effective gluon
correlator models  the interaction between quark currents.  Functional integral
calculus allows the GCM to be hadronised (i.e. expressed in terms of meson and baryon fields). The  dominant
configuration of the hadronic functional integrals is revealed to be the constituent quark effect, and 
  is identical to   one version of the truncated quark Dyson-Schwinger 
equations (tDSE). However the GCM shows that hadronic physics requires processes
that go beyond the  tDSE. In this review examples of meson and nucleon processes are given.
The GCM also plays a pivotal role  in  showing how QCD may be related to 
many other hadronic models}

\vspace{1cm}

\vspace{0.3cm}


\end{minipage} \end{center}

\newpage

\tableofcontents

\section{Introduction}

\begin{figure}[ht]
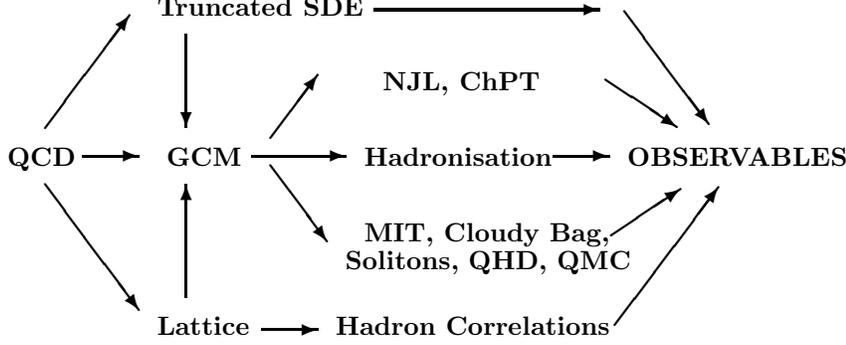
 
\vspace{-15mm}\caption{\small{Relational map of the GCM to QCD and various other
modellings including the Nambu - Jona-Lasinio (NJL), Quantum Hadrodynamics (QHD),
Quark Meson Coupling (QMC)  and Chiral Perturbation Theory (ChPT).}\label{figure:Map}} 
\end{figure}

We review the Global Colour Model (GCM) (Cahill and
Roberts (1985)\cite{CR85}) of Quantum Chromodynamics (QCD) with particular emphasis on its
hadronisation and the resulting applications to low energy meson and nucleon processes.
Other reviews of the GCM are Cahill (1989)\cite{C89a}, Cahill (1992)\cite{C92}, 
Roberts and Williams (1994)\cite{RW}, Tandy (1996)\cite{Tand96}, Cahill and Gunner
(1997)\cite{CG97a} and Tandy (1997)\cite{Tandy}. 
 QCD is defined
by the quantisation of the  quark-gluon  fields  with `classical' action
$S[\overline{q},q,A^a_{\mu}]$. However all
evidence for the quarks and gluons is provided by the properties and interactions of the
hadrons  and   by processes involving the electroweak particles. These hadronic laws are
encoded in an effective action $S[\pi,\rho,\omega,..,\overline{N},  N,.]$,  where
$\pi(x),..,\overline{N}(x),N(x),...$ are fields describing composite constituent 
(equivalently core or bare) hadrons, with
`centre-of-mass coordinates' $x$.  These hadronic fields are to be quantised subject to this
effective action,  yielding finally the observables of QCD. Such an effective action must clearly be
non-local because of the composite nature of these hadrons. While a general derivation of 
  $S[\pi,..,\overline{N},N,.]$ from  $S[\overline{q},q,A^a_{\mu}]$ has not  been
achieved, it is possible to do this hadronisation  within the  GCM.
The hadronisation uses Functional Integral Calculus (FIC) techniques, which amount to 
dynamically determined changes of integration variables   in the functional integral 
formulation  of the GCM; 
\begin{equation}                                                                  \label{eqn;A1}               
\int D{\overline q}DqDA\exp(-S_{GCM}[A,{\overline q},q])                     
   =\int  D\pi ..D{\overline N}DN..\exp(-S_{had}[\pi,..,{\overline N},N,..]).          
\end{equation} 
Here the constituent hadrons are essentially the normal modes. 
A particular feature of the GCM is that it plays a pivotal role in relating various seemingly
different modellings of QCD as shown in Fig.\ref{figure:Map}, and these relationships will be
discussed later. A key task in using the GCM is the determination of the low energy quark-gluon processes from
experimental data.  In recent years there have been three such extractions of increasing complexity:
GCM95\cite{CG95a}, GCM97\cite{CG97b} and GCM98\cite{CG98a}.  We report here  new detailed {\it ab
initio} studies of the nucleon within the GCM in which one proceeds in a rigorous manner from the
experimentally determined low energy quark-gluon processes to detailed dynamical studies of the
nucleon, including dressing by the pions. 

\vspace{5mm}
\setlength{\unitlength}{0.25mm}
\begin{center}
\hspace{-30mm}
\begin{picture}(10,180)(+200,80)
\thicklines
\put(5,200){\bf QCD}
\put(25,190){\vector(3,-4){50}}
\put(45,205){\vector(3,0){30}}
\put(85,200){ \bf GCM}
\put(145,215){\vector(3,4){25}}
\put(200,240){ \bf NJL, ChPT}
\put(135,205){\vector(3,0){50}}
\put(190,200){ \bf Hadronisation}
\put(295,205){\vector(3,0){30}}

\put(333,282){\vector(3,-4){45}}
\put(323,246){\vector(3,-2){38}}
\put(326,163){\vector(3,+2){37}}
\put(328,115){\vector(3,+4){55}}

\put(330,200){ \bf OBSERVABLES}
\put(145,200){\vector(3,-4){30}}
\put(190,160){ \bf MIT, Cloudy Bag,  }
\put(180,145){ \bf Solitons, QHD, QMC}
\put(80,110){ \bf Lattice }
\put(140,113){\vector(1,0){30}}
\put(100,130){\vector(0,1){60}}
\put(175,110){ \bf Hadron Correlations}
\put(80,280){ \bf Truncated SDE }
\put(200,283){\vector(1,0){120}}
\put(25,220){\vector(3,+4){45}}
\put(100,270){\vector(0,-1){50}}

\end{picture}
\end{center}

\section{Global Colour Model\label{section:GlobalColourModel}}

\begin{figure}[ht]
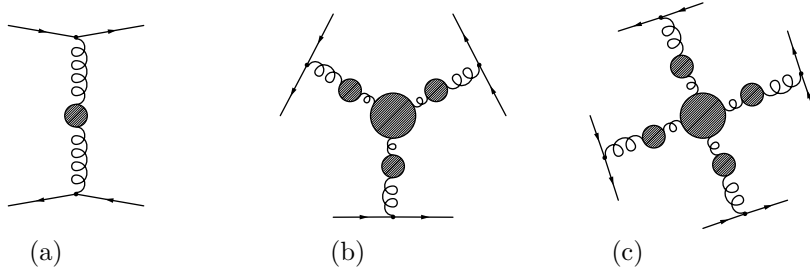

\vspace{-7mm}
\caption{\small{Diagrammatic representation of successive terms in the quark action in 
(\ref{eqn;R3}).  The quark-gluon vertex strength is $g_0$, while the
gluon-gluon vertices (including gluon correlators)  are fully dressed except for
quark loops.}
 \label{figure:QCDaction}}
\end{figure}

\begin{figure}[ht]
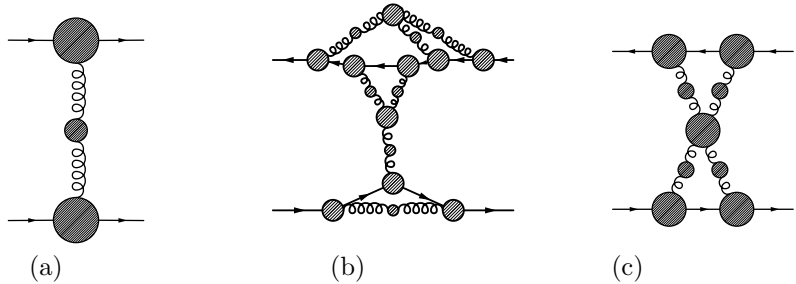

\vspace{-5mm}\caption{\small{(a) The GCM effective $D_{\mu\nu}$ in
(\ref{eqn;R5}), (b) example of correlations formally included  in (a), and in (c)  an $n=4$ process not
formally included in (a), but which is modelled in the GCM  via the specific form of $D_{\mu\nu}$.}
\label{figure:Ints} }
\end{figure}

 Here we discuss the construction of the GCM from QCD.  
In the functional integral approach correlators are defined by 
\begin{equation}                                                                            \label{eqn;R1}
{\cal G}(..,x,...)=\frac{\int{D}\overline{q}{D}q{DA}{D}\overline{C}{D}C....q(x).....
\mbox{exp}(-S_{QCD}[A,\overline{q},q,\overline{C},C])} {\int{D}\overline{q}{D}q{DA}{D}\overline{C}{D}C
\mbox{exp}(-S_{QCD}[A,\overline{q},q,\overline{C},C])}
\end{equation}
where the  `classical' action defining chromodynamics  is, in Euclidean metric,
\begin{eqnarray}                                                                                 \label{eqn;A2}      
S_{QCD}[\overline{q},q,A^a_{\mu}]=\int d^4x{\biggr (}\frac{1}{4}F^a_{\mu\nu}F^a
_{\mu\nu}&+&\frac{1}{2\xi}(\partial_{\mu}A^a_{\mu})^2  \nonumber \\ 
&+&\overline{q}(\gamma
_{\mu}(\partial_{\mu} -ig\frac{\lambda^a}{2}A^a_{\mu})+{\cal M})q{\biggr )}.
\end{eqnarray}
This involves the quark and gluon  fields and the field strength tensor for the gluon
fields.  ${\cal M}=\{m_u,m_d,..\}$ are the quark current masses, and ghost ($\overline{C},C$) and 
gauge fixing terms must be added to $S_{QCD}$ in (\ref{eqn;A2}). The chromodynamic action  
clearly has two important invariance groups, Poincar\'{e}
 symmetry and the local colour symmetry. 
The  various complete (denoted by scripted symbols) correlators ${\cal G}$ lead to experimental
observables. They  are related by an infinite set of coupled Dyson-Schwinger Equations (DSE), and
by the Slavnov-Taylor gauge-symmetry identities and,  in the chiral limit, to the axial
Ward-Takahashi identity (AWTI).  The usual truncation of these DSE causes the violation of all
these identities.
 The  correlators in (\ref{eqn;R1}) may be extracted from the generating functional of QCD, 
\begin{equation}                                                                            \label{eqn;R2}
Z_{QCD}[\overline{\eta},\eta,J]=\int { D}\overline{q}{ D}q{
D}A{ D}\overline{C}{
D}C\mbox{exp}(-S_{QCD}[A,\overline{q},q,\overline{C},C]+\overline{\eta}q+
\overline{q}\eta+JA).
\end{equation}
Functional transformations and approximations lead to the GCM; briefly  
and not showing source terms for convenience,  the gluon and ghost integrations are formally
performed
\begin{eqnarray}                                                          \label{eqn;R3}
\int { D}\overline{q}{ D}q{ D}A{D}\overline{C}{D}C\mbox{exp}(-S_{QCD}) 
= \int {D}\overline{q}{D}q\mbox{exp}(-\int 
\overline{q}(\gamma . \partial+{\cal M})q  \nonumber \\ 
+\frac{g_0^2}{2}\int j^a_{\mu}(x)j^a_{\nu}(y){\cal G}_{\mu\nu}(x-y)+\frac{g_0^3}{3!}
\int j^a_{\mu}j^b_{\nu}j^c_{\rho}{\cal G}^{abc}_{\mu\nu\rho}+......),
\end{eqnarray}
where $j^a_{\mu}(x)=\overline{q}(x)\frac{\lambda^a}{2}\gamma_{\mu}q(x)$, 
$g_0$ is the bare coupling
constant, and ${\cal G}_{\mu\nu}(x)$ is  the  gluon
correlator  with no quark loops but including ghosts ($\overline{C},C$)
\begin{equation}                                                            \label{eqn;R4}
{\cal G}_{\mu\nu}(x-y)=
\frac{\int {D}A{\cal D}\overline{C}{D}CA^a_{\mu}(x)A^a_{\nu}(y)
\mbox{exp}(-S_{QCD}[A,\overline{C},C])}{\int {D}A{D}\overline{C}{D}C
\mbox{exp}(-S_{QCD}[A,\overline{C},C])}.
\end{equation}  
Fig.\ref{figure:QCDaction} shows successive terms in (\ref{eqn;R3}). This infinite sequence is a direct
consequence of the local non-abelian colour symmetry. The terms of higher order than the term quartic in the
quark fields   are difficult to explicitly  retain in any analysis.  The GCM models the effect of 
higher order terms by replacing the coupling constant
$g_0$ by a momentum dependent quark-gluon coupling $g(s)$, and neglecting terms like ${\cal
G}^{abc}_{\mu\nu\rho}$  and higher order in (\ref{eqn;R3}). This $g(s)$ is a restricted form of vertex
function.  The modification 
$g_0^2{\cal G}_{\mu\nu}(p) \rightarrow D_{\mu\nu}(p) = g(p^2){\cal G}_{\mu\nu}(p)g(p^2)$
and the truncation  then defines the GCM.  We also call this effective quark-quark coupling
correlator $D_{\mu\nu}(p)$ the effective gluon correlator and show included processes in Fig.\ref{figure:Ints}.
 As discussed here $D_{\mu\nu}(p)$ may be determined from experimental data.
We then formally define the  GCM as the quark-gluon field theory 
with the action
\begin{equation}                                                                 \label{eqn;R5}
S_{GCM}[\overline{q},q,A^a_{\mu}]=\int \left( 
\overline{q}(\gamma . \partial+{\cal M}-iA^a_{\mu}\frac{\lambda^a}{2}\gamma_{\mu})q
 +\frac{1}{2} A^a_{\mu}D^{-1}_{\mu\nu}(i\partial)A^a_{\nu} \right)
\end{equation} 
and the generating functional
\begin{equation}                                                                \label{eqn;A3}      
Z[J,\overline{\eta},\eta]=\int { D}\overline{q}{ D}q{ D}A
\exp(-S_{GCM}[\overline{q},q,A^a_{\mu}]+
\overline{\eta}q+\overline{q}\eta+J^a_{\mu}A^a_{\mu}). 
\end{equation}

\begin{minipage}[t]{35mm}
\hspace{5mm}\includegraphics[scale=0.3]{Fig1a.EPSF}

 \makebox[20mm][c]{(a)}
\end{minipage}
\begin{minipage}[t]{40mm}  
\hspace{5mm}\includegraphics[scale=0.3]{Fig1b.EPSF}
\makebox[28mm][c]{(b)}
\end{minipage}
\begin{minipage}[t]{40mm}
 \hspace{5mm}\includegraphics[scale=0.3]{Fig1c.EPSF}
\makebox[20mm][c]{(c)}
\end{minipage}

\begin{minipage}[t]{35mm}
\hspace{5mm}\includegraphics[scale=0.3]{Fig2a.EPSF} 
 \makebox[20mm][c]{(a)}
\end{minipage}
\begin{minipage}[t]{40mm}
\vspace{-30mm}  
\hspace{0mm}\includegraphics[scale=0.4, bb =0 +40 400 250]{Fig2b.EPSF}
\makebox[28mm][c]{(b)}
\end{minipage}
\begin{minipage}[t]{40mm}
 \hspace{5mm}\includegraphics[scale=0.3]{Fig2c.EPSF}
\makebox[20mm][c]{(c)}
\end{minipage}

Here $D_{\mu\nu}^{-1}(p)$ is the matrix inverse of the Fourier transform of $ D_{\mu\nu}(x)$.  
The action $S_{GCM}$ is invariant under $q\rightarrow Uq,\mbox{ }
\overline{q}\rightarrow \overline{q}U^\dagger$ and $A^a_{\mu}\lambda^a \rightarrow U
A^a_{\mu}\lambda^a U^\dagger$,  where $U$ is a global $3\times3$ unitary colour matrix; this is the
global colour symmetry of the GCM.
 The gluon  self-interactions  that
arise as a consequence of  the local colour symmetry in (\ref{eqn;R4}) and the ghost and vertex
effects   lead to 
$D_{\mu\nu}^{-1}(p)$ in (\ref{eqn;R5}) being  non-quadratic. Hence, in effect, the GCM models the QCD local
gluonic action  $\int F^a_{\mu\nu}[A]F^a_{\mu\nu}[A]$ in
$S_{QCD}$ of (\ref{eqn;R1}) which has local colour symmetry, by a highly
nonlocal action in the last term of (\ref{eqn;R5}) which has global colour symmetry. It is
important to appreciate that while the GCM has a formal global colour symmetry, the detailed dynamical
consequences of the local colour symmetry of QCD are modelled by the particular form of $D(s)$. 
There is an Infrared (IR) saturation effect which, in conjunction with the dynamical breaking of chiral
symmetry,  appears to suppress details of the formal colour gauge symmetry of QCD.
 As well, in the chiral limit ${\cal M} \rightarrow 0$, the GCM action has
 $U_L(N_F)\otimes U_R(N_F)$  symmetry:  the $\overline {q}-q$
 part of $S$ may  be written $\overline{q}\gamma_{\mu}q=\overline{q}_R\gamma_{\mu}q_R+
\overline{q}_L\gamma_{\mu}q_L$, where $q_{R,L}=P_{R,L}q$ and
$\overline{q}_{R,L}=\overline{q}P_{L,R}$. These two   parts are separately
invariant under $q_R\rightarrow U_Rq_R,\mbox{ }
\overline{q}_R\rightarrow\overline{q}_RU^{\dagger}_R$ \mbox{ }  and $q_L\rightarrow U_Lq_L, \mbox{ } 
\overline{q}_L\rightarrow\overline{q}_LU^{\dagger}_L$.

A key feature of the GCM analysis is the demonstration that this effective gluon correlator $D_{\mu\nu}(p)$
is successful for a variety of hadronic processes, i.e. it is a {\it universal} feature of low
energy hadronic processes. The success of the GCM appears to be based on the phenomenon of an IR saturation
mechanism in which the extreme IR strength of the many contributing  quark-quark couplings  is
easily modelled by this one effective gluon correlator.   This was used in the context
of the truncated Dyson-Schwinger (tDSE) approach by Munczek and Nemirovsky \cite{MN83} who used a
delta-function form. Of particular dynamical interest  is the comparison 
of the GCM $D_{\mu\nu}(p)$ with one constructed theoretically from only a 
gluon correlator and vertex functions, say from continuum or lattice modellings, for this  
gives some insight into the IR strength of the higher order gluonic couplings.

\section{Hadronisation\label{section:Hadronisation}}

Hadronisation is a generalisation of the bosonisation concept to include the 
fermionic (baryonic) states. Bosonisation is   naturally and conveniently induced using the FIC methods,
and indeed the GCM is a spinor-boson field theory which may be exactly bosonised in 4-dimensional
spacetime.  A key feature of the GCM bosonisation is
the use of bilocal fields which simply indicates that the complete theory is expressible using  
two-point correlators as an equivalent set of functional integration variables.   As we shall see the
bilocal bosonisation (which precedes the hadronisation) not only generates the
 mesonic  effective action, but also carries structural information which is essential in  understanding the
couplings of these mesons. This information (i.e. vertex functions)  is essential to the prediction of hadronic
properties and interactions. If we throw away this emergent information we are left with a non-renormalisable
effective field theory with no predictive properties.    
There are two alternative  first stage bosonisations of the GCM. The first in 1985 in which Cahill and 
Roberts\cite{CR85}  introduced colour singlet and colour octet mesonic correlations had two problems: (a) the
colour octet correlations did not appear to have any clear  physical significance, and (b)  the baryonic states were
not manifest.  However in 1989 Cahill,  Praschifka and Burden\cite{CPB89} discovered a meson-diquark bosonisation  
which was induced by  new colour and Dirac-spinor Fierz identities and which involved the colour singlet
mesonic states and the colour ${\bf\overline{3}}$ ($qq$) and ${\bf 3}$   ($\overline{q}\overline{q}$)
diquark/anti-diquark states, these being the very colour subcorrelations present in colour singlet baryons and
anti-baryons.  The meson-diquark
bosonisation then led to the meson-baryon hadronisation of the GCM (Cahill 1989)
\cite{C89b}. This hadronisation automatically produced the covariant Faddeev formulation of the
constituent baryon  states, and also the meson dressing of these states; see later sections.
In this review we concentrate mainly on this modern hadronisation of the GCM, which is summarised by the
following sequence of FIC transformations:
\begin{eqnarray}                                                     
Z&=&\int {D}\overline{q}{D}q{D}A\exp(-S_{QCD}[A,\overline{q},q]+\overline{\eta}q
+\overline{q}\eta) \nonumber \\
&\approx&\int {D}\overline{q}{D}q{D}A\exp(-S_{GCM}[A,\overline{q},q]+\overline{\eta}q+
\overline{q}\eta)   \mbox{\ \ \ \ (GCM)} \nonumber \\
 &=&\int {D}{\cal  B}{D}\overline{\cal D}D{\cal D}\mbox{exp}(-S[
{\cal B},\overline{\cal D},\cal D]) \mbox{\ \ \ \ (bilocal fields)}  \label{eqn;R6}  \\ 
&=&\int{D}\overline{N}{D}N..{D}\pi{D}\rho{D}\omega...
\mbox{exp}(-S_{had}[\overline{N},N,..\pi,\rho,\omega....])  \mbox{\ \ (local fields). } \label{eqn;R7} 
\end{eqnarray} 

The basic insights are that (i) the quark-gluon dynamics in (\ref{eqn;R1})  is
fluctuation dominated, whereas the hadronic functional integrations in  (\ref{eqn;R7}) are
not, (ii) the bilocal stage in (\ref{eqn;R6}) produces the constituent quark effect as the dominant
configuration, and (iii) this entails the IR saturation effect and the dynamical breaking of chiral symmetry
and its significant consequences, and (iv) the  induced  hadronic effective action in (\ref{eqn;R7}) is
nonlocal.  
The hadronisation has also been further studied in 
\cite{Reinhardt90,Ebert:1991qg,Ebert:1997hr,Ebert:1992zq,Ebert:1993xd,Ebert:1997ey}

\subsection{Meson-Diquark Bosonisation}

Here we review the meson-diquark bosonisation \cite{C89a} of the GCM, giving
(\ref{eqn;R6}). The FIC techniques amount to analogues
of various `tricks' of ordinary integral calculus.  Integrating out the gluon fields
$Z$ becomes,
\begin{equation}                                                               \label{eqn;A10}      
 Z[\overline{\eta},\eta]=
\int {D}\overline{q}{D}q\exp\Bigl(-S[\overline{q},q]+\overline{ \eta}q
+\overline{q}\eta\Bigr),
\end{equation}
where
\begin{eqnarray}                                                            \label{eqn;A11}      
S[\overline{q},q]&=& \int \Bigl(
\overline{q}(x)(\gamma . \partial_x+{\cal M}) \delta^4(x-y)q(y) +\frac{1}{2}\overline{q}(x)\frac{\lambda^a}{2}\gamma_{\mu}
q(x)D^{ab}_{\mu\nu}(x-y)\overline{q}(y)\frac{\lambda^b}{2}\gamma_{\nu}q(y)\Bigr). 
\end{eqnarray} 
Using the new Fierz identities \cite{CPB89} the quartic term in
(\ref{eqn;A11}) is rearranged to give 
\begin{eqnarray}                                                                   \label{eqn;A12}       
S[\overline{q},q]&=& \int d^4xd^4y\Bigl[\overline{q}(x)\gamma.\partial\delta^4(x-y)q(y)
     -\frac{1}{2}\overline{q}(x)\frac{M^{\theta}_m}{2}q(y)D(x-y).  \nonumber \\ 
& &.\overline{q}(y)\frac{M^{\theta}_m}{2}q(x)
 -\frac{1}{2}\overline{q}(x)\frac{M^{\phi}_d}{2}\overline{q}(y)
  ^{cT}D(x-y)q(y)^{cT}\frac{M^{\phi}_d}{2}q(x)\Bigr],
\end{eqnarray}
with $q^c =Cq$ , $\overline{q}^c=\overline{q}C$. 
The Fierz identities are the two Dirac matrix identities
$\gamma^{\mu}_{rs}\gamma^{\mu}_{tu}=K^a_{ru}K^a_{ts}$ where 
  $\{K^a\}=\{{\bf 1}, i\gamma_5, \frac{i}{\surd 2}\gamma^{\mu},
  \frac{1}{\surd 2}\gamma^{\mu}\gamma_5\},$ and 
$\gamma^{\mu}_{rs}\gamma^{\mu}_{tu}=(K^aC^T)_{rt}(C^TK^a)_{us}$
 where $C=\gamma^2\gamma^4,$  $C^2=-{\bf 1}$ and $C\gamma^{\mu}C= \gamma^{\mu T}.$ 
\noindent We also  use the set $\{\overline{K}^a\}=\{{\bf 1},
-i\gamma_5, \frac{-i}{2\surd 2}\gamma^{\mu},
  \frac{1}{2\surd 2}\gamma^{\mu}\gamma_5\}$, then
$tr[\overline{K}^aK^b]=4\delta_{ab}$.  
For  the colour algebra \cite{CPB89}
$\lambda^a_{\alpha\beta}\lambda^a_{\gamma\delta}=\frac{4}{3}
\delta_{\alpha\delta}\delta_{\beta\gamma}+\frac{2}{3}\sum_{\rho=1}^{3}
\epsilon_{\rho\alpha\gamma}\epsilon_{\rho\delta\beta},$ 
while for the $N_f=3$ flavour algebra, 
$\delta_{ij}\delta_{kl}=F^c_{il}F^c_{kj}$ for the mesons where 
$\{F^c,c=0,..8\}= \{\frac{1}{\surd 3}{\bf 1},\frac{\lambda^1}{\surd 2},..,
\frac{\lambda^8}{\surd 2}\}$  and
$\delta_{ij}\delta_{kl}=H^f_{ik}H^f_{lj}$ for the diquarks, where $ \{H^f, f=1,..9\}=\{F^c, 
c=7,5,2, 0,1,3,4,6,8\}$ and
where $\{\frac{\lambda^a}{2}\}$ are the generators of $SU(3)$ in the usual Gell-Mann representation.
We  define the tensor products $\{M^\theta_m\}=\{\surd\frac{4}{3}K^aF^c\}$
and $\{M^{\phi}_d\}=\{i\surd\frac{2}{3}K^a\epsilon^{\rho}H^f\}$, where
$(\epsilon^{\rho})_{\alpha\beta} = \epsilon_{\rho\alpha\beta}$. We see that
$\overline{q}(y)M^{\theta}_mq(x)$ are ${\bf 1}_c$ bilocal $\overline{q}q$ fields with the
flavour (${\bf 1}_f$ or ${\bf 8}_f$) 
determined by the  flavour generators  ($\{F^0\}$ or $ \{F^{1,..,8}\}$) in $M^{\theta}_m$,
 while $q(y)^{cT}M^{\phi}_dq(x)$
are $\overline{{\bf 3}}_c$  bilocal $qq$ fields with the flavour
($\overline{{\bf 3}}_f$ or ${\bf 6}_f$) determined by the flavour
generators  ($\{H^{1,2,3}\}$ or $\{H^{4,..,9}\}$) in $M^{\phi}_d$. These
results follow from the colour and flavour representation of the quark fields. The (integral)
spin of these boson fields is determined by the $K^a$. By rewriting 
(\ref{eqn;A11}) as (\ref{eqn;A12}) we can initiate  a bosonisation which is adapted to 
the attractive channels implicit in (\ref{eqn;A11}).  The 1985 GCM colour Fierz identity \cite{CR85}
 leads to the colour ${\bf 8}$ channels which are repulsive.

We make the first FIC change of variables by noting that the
quartic terms in $\exp(-S)$ may be generated by the following bilocal FIC
integrations, 
\begin{eqnarray}                                                            \label{eqn;A13}      
Z&=&\hspace{-1em}\int D\overline{q}Dq D{\cal B}D{\cal D}D{\cal D}^{\star}\exp \Bigl( \int
   \Bigl[ -\overline{q}(x)(\gamma.\partial+{\cal M})\delta^4(x-y)q(y)  \nonumber \\ 
  & &-\frac{{\cal B}^{\theta}(x,y){\cal B}^{\theta}(y,x)}{2D(x-y)} 
  -\frac{{\cal D}^{\phi}(x,y){\cal D}^{\phi}(x,y)^{\star}}{2D(x-y)}  \nonumber \\ 
  & &-\overline{q}(x)\frac{M^{\theta}_m}{2}q(y){\cal B}^{\theta}(x,y) 
 -\frac{1}{2}\overline{q}(x)\frac{M^{\phi}_d}{2} 
\overline{q}(y)^{cT}{\cal D}^{\phi}(x,y)^{\star}  \nonumber \\ 
& &-\frac{1}{2}{\cal D}^{\phi}(x,y)q(y)^{cT}
\frac{M^{\phi}_d}{2}q(x)\Bigr]+ \int (\overline{\eta}q+\overline{q}\eta)\Bigr), 
\end{eqnarray}
where ${\cal B}^{\theta}(x,y)={\cal B}^{\theta}(y,x)^{\star}$ are
`hermitean' bilocal fields.
Integration over the quark fields completes the change of variables
to bilocal meson and diquark fields,  
\begin{eqnarray}                                                                 \label{eqn;A14}      
Z[\overline{\eta},\eta]&=&\hspace{-1em}\int D{\cal B}D{\cal D}D{\cal D}^{\star}
(Det{\cal F}^{-1}[{\cal B},{\cal D},\overline{{\cal D}}])^{\frac{1}{2}}
.\exp{\biggr ( }\int-\frac{{\cal B}^{\theta}(x,y) {\cal B}^{\theta}(y,x)}{2D(x-y)} \nonumber \\
&&\mbox{\ \ \ \ \ \ \ \ \ \  }-\int  \frac{{\cal D}^{\phi}(x,y){\cal D}^{\phi}(x,y)^{\star}}{2D(x-y)}
+\frac{1}{2}\int \Theta{\cal F}\Theta^T{\biggr )},
\end{eqnarray}
where $\Theta\equiv(\overline{\eta},-\eta^T),$
$\mbox{\ \ \ \ \ }{\cal F}^{-1}[{\cal B}, {\cal D},
\overline{{\cal D}}]= \left(\begin{array}{cc} -{\cal D} & G^{-1T} \\
-G^{-1} & -\overline{{\cal D}} \end{array} \right),$
\begin{equation}                                                                \label{eqn;A15}
G^{-1}(x,y,{\cal
B})=(\gamma.\partial+{\cal M})\delta^4(x-y)+{\cal B}(x,y), \mbox{\ \ \ \ }  {\cal B}(x,y)={\cal
B}^{\theta}(x,y)\frac{M^{\theta}_m}{2},
\end{equation}                                                              
   \[  \mbox{\ \ \ \ }\overline{\cal D}(x,y)={\cal
D}^{\phi}(x,y)^{\star} \frac{M^{\phi}_d}{2}C^T \mbox{\ \ \ \ \ \ and \ \ \ \ }
{\cal D}(x,y) = {\cal D}^{\phi}(y,x)C^T\frac{M^{\phi}_d}{2}.\] \noindent
Using the determinant identity \cite{CPB89}
$Det{\cal F}^{-1}=(Det(G^{-1}))^2Det({\bf 1}+
\overline{\cal D}G^T{\cal D}G), $
\begin{eqnarray}                                                            \label{eqn;A16}      
Z=\int D{\cal B}D{\cal D}D{\cal D}^{\star}
   \exp\left(TrLn(G[{\cal B}]^{-1})+\frac{1}{2}TrLn({\bf 1}+
\overline{\cal D}G[{\cal B}]^T{\cal D}G[{\cal B}])+\right.  \nonumber \\
\mbox{\ \ \ \ \ \ \ \ \ \ \ \ \ \ \ \ \ \ \ \ \ \ \ \ \ \ \ \ \ \ \ \ \ \ \ }\left.-\int
\frac{{\cal B}^{\theta}{\cal B}^{\theta\star}}{2D}- \int \frac{{\cal D}^{\phi}{\cal
D}^{\phi\star}}{2D}+\frac{1}{2}\int \Theta{\cal F}\Theta^T\right).
\end{eqnarray}

\subsection {Baryons}

\begin{figure}[ht]
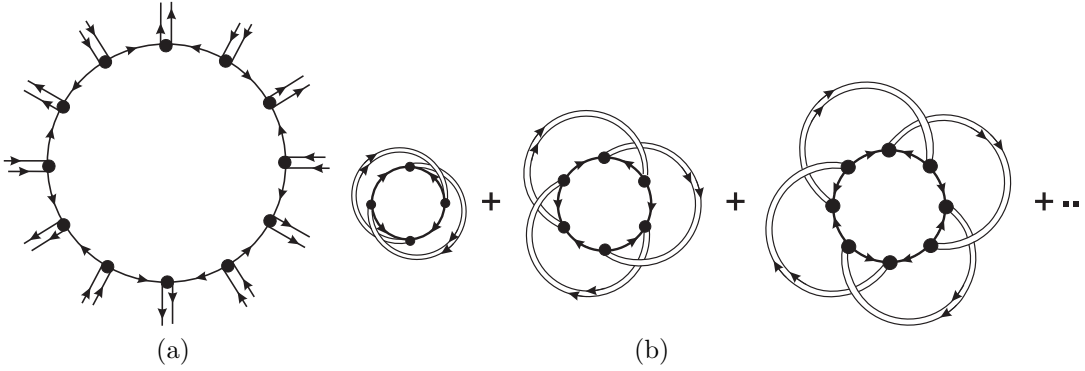

\vspace{-5mm}\caption{\small{ (a) Example of diagrams from expansion of diquark $TrLn$ in (\ref{eqn;A17}).
 (b) Diagrams  after diquark field integrations.}
 \label{figure:loop}}
\end{figure}

  The diquark sector of the meson-diquark bosonisation
 generates \cite{C89a} the colour singlet baryon states of the GCM.  Consider the diquark part of
Z;\[Z=\int D{\cal D}D{\cal D}^{\star}\exp\left(\frac{1}{2} TrLn({\bf 1}+\overline{\cal D}G^T{\cal
D}G)-\int \frac{{\cal D} {\cal D}^{\star}}{2D}
+\int (J^{\star}{\cal D}+{\cal D}^{\star}J)\right),\] \noindent
where the bilocal diquark source terms facilitate the analysis, and
in which the ${\cal B}$  dependence of $G[{\cal B}]$ will  affect both the non-trivial
dominant configuration and the mesons, and  will provide the meson-baryon couplings.  The expansion 
\begin{equation}                                                           \label{eqn;A17}      
TrLn({\bf 1}+\overline{\cal D}G^T{\cal D}G)=\sum_{n=1}^{\infty}\frac{(-1)^{(n+1)}}{n}
               Tr(\overline{\cal D}G^T{\cal D}G)^n,
\end{equation}
gives single loop processes (Fig.\ref{figure:loop}(a)) with the quark lines
alternating in direction, in accord with their coupling to the diquark
and anti-diquark fields. 
Using (\ref{eqn;A17}) the  diquark part of the action  has the expansion
$S[{\cal D}^{\star}, {\cal D}]=\sum_n S_n[{\cal D}^{\star}, {\cal D}]$ and 
with $S_1=\int  {\cal D}^{\phi\star}(\Delta_d^{-1})^{\phi\psi}
{\cal D}^{\psi}$
and $S^{\prime}=S-S_1$, 
\begin{eqnarray}                                                              \label{eqn;A18}      
Z&=&\exp(-S^{\prime}[\frac{\delta}{\delta J^{\star}},
\frac{\delta }{\delta J}])\int D{\cal D}D{\cal D}^{\star}
\exp\left(-\int{\cal D}^{\star}\Delta_d^{-1}{\cal D}+\int(J^{\star}{\cal D}
+{\cal D}^{\star}J)\right) \nonumber \\
&=&\exp(-S^{\prime}[\frac{\delta }{\delta J^{\star}},
\frac{\delta }{\delta J}])\exp\left(-TrLn(\Delta_d^{-1})+\int J^{\star}
\Delta_dJ\right). 
\end{eqnarray}

 
\hspace{0mm}
\begin{minipage}[t]{50mm}
\hspace{-5mm}\includegraphics[scale=0.6]{loop.EPSF}
\makebox[35mm][c]{(a)}
\end{minipage}
\begin{minipage}[t]{50mm}
\hspace{-10mm}\includegraphics[scale=0.6]{diquark_loop.EPSF}
\makebox[60mm][c]{(b)}
\end{minipage}

\begin{figure}[ht]
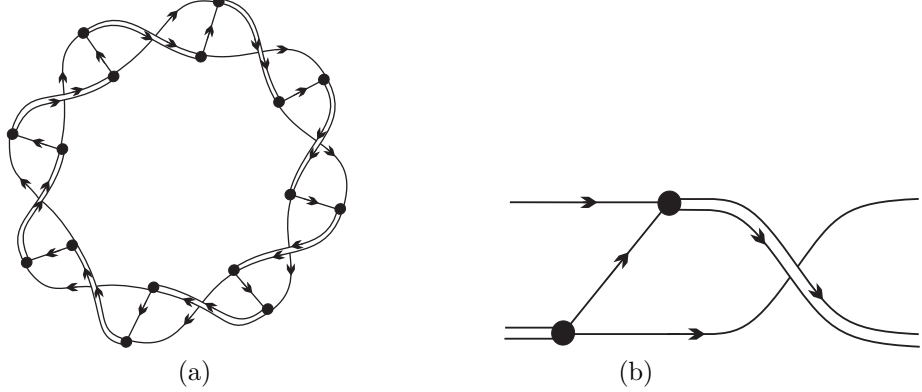

\vspace{-5mm}\caption{\small{ (a) Similar to  Fig.\ref{figure:loop}(b) after redrawing to reveal
baryon loop. These loop functionals determine the core baryon mass spectrum. (b) Kernel of the nucleon
loop.}
 \label{figure:ConstituentQuarkI}}
\end{figure}

\noindent Keeping only the translation invariant part of ${\cal B}$,  ${\cal B}_{CQ}(x-y)$ - later to be
identified  as the constituent quark (CQ) effect, $\Delta_d^{-1}$ has eigenvalues and eigenvectors (diquark
form factors) from
\begin{equation}                                                             \label{eqn;A19}      
\int \frac{d^4q}{(2\pi)^4}(\Delta_d^{-1})^{\phi\psi}(p,q;P)
\Gamma_k^{\psi}(q;P)=  \lambda_k(P^2)\Gamma_k^{\phi}(p;P) 
\end{equation} 
and we have the orthonormality,  completeness  and  spectral equations
\[\sum_{\theta}\int\frac{d^4q}{(2\pi)^4}\Gamma^{\theta}_k(q;P)^*
\Gamma^{\theta}_l(q;P)=\delta_{kl},\mbox{\ }
\sum_k\Gamma^{\theta}_k(q;P)\Gamma^{\phi}_k(p;P)^*=
(2\pi)^4\delta_{\theta\phi}\delta^4(q-p),\] 
\begin{equation}                                                               \label{eqn;A20}      
\Delta_d^{\phi\psi}(p,q;P)=\sum_k\Gamma_k^{\phi}(p;P)\lambda_k(P^2)^{-1}
\Gamma_k^{\psi}(q;P)^*.
\end{equation}
Using completeness  we construct the local-diquark-field FIC representation
\begin{equation}                                                                \label{eqn;A21}   
TrLn(\Delta^{-1}_d)=\sum_k\int
d^4x\int\frac{d^4P}{(2\pi)^4}ln(\lambda_k(P^2))
                 =\sum_kTrLn(\lambda_k({-\partial^2})\delta^4(x-y)), 
\end{equation}
\begin{equation}                                                                \label{eqn;A22}   
\exp(-TrLn(\Delta_d^{-1}))=\int Dd_kDd^\star_k\exp\Bigl(-\sum_k \int
d_k(x)^\star\lambda_k({-\partial^2})\delta^4(x-y)d_k(y)\Bigr).
\end{equation}
Introducing local sources $j_k(X)=\int d^4Yd^4x\Gamma_k^{\phi}(x,X-Y)^*J^{\phi}(x,Y)$,
so that $$\frac{\delta}{\delta J^{\phi}(x,X)}=\sum_k\int
d^4Y\Gamma_k^{\phi}(x,Y-X)^*\frac{\delta}{\delta j_k(Y)},$$  but  keeping only a single component
of the scalar diquark to simplify notation, 
\begin{equation}                                                              \label{eqn;A24}   
Z[j^{\star},j]=\exp(-S^{\prime}[\frac{\delta}{\delta
j^{\star}},\frac{\delta}{\delta  j}]) \exp\left(-TrLn(\Delta^{-1}_d)+\int
j^{\star}(X)\lambda_0({-\partial^2})^{-1}j(X)\right). 
\end{equation}
Evaluating  the effect of the
functional operator we find that $Z[0,0]$ has the form $\exp(W)$ where $W$ is the sum of
connected loop diagrams,
with the vertices  now joined by the diquark correlators
$\lambda_0(P^2)^{-1}=(P^2+m_0(P^2)^2)f^2$, in which  $m_0(P^2)$ is the diquark mass
function, and  with $\Gamma_0(p;P)$ at the vertices.  
 Of particular significance is the infinite
subset of diagrams which will be seen to have the form of  three-quark (i.e. baryon) loops
(Fig.\ref{figure:ConstituentQuarkI}(a)). These come with a combinatoric factor of 2
(except for the order n=3 diagram)
which cancels the $\frac{1}{2}$ coefficient of the $TrLn$.
  These 3-loops are planar for even order, but non-planar,
with one twist, for odd order. To exhibit the three-quark loop structure
 we show, in Fig.\ref{figure:ConstituentQuarkI}(a), a typical diagram from 
Fig.\ref{figure:loop}(b) after  deformation,
revealing a closed double helix: a diagram of order $n$ is drawn on a
M\"{o}bius strip of $n-1$ twists.
This infinite series may be summed  as the diagrams are generated by the
kernel $K$, defined by the one-twist diagram, shown in Fig.\ref{figure:ConstituentQuarkI}(b).

\hspace{0mm}\begin{minipage}[t]{50mm}
\hspace{0mm}\includegraphics[scale=1.0]{nucleon_loop.EPSF}
\makebox[50mm][c]{(a)}
\end{minipage}
\begin{minipage}[t]{50mm}
\hspace{15mm}\includegraphics[scale=1.4]{nucleon_kernel.EPSF}
\makebox[65mm][c]{(b)}
\end{minipage}

\begin{figure}[ht] 
\hspace{20mm}\includegraphics[scale=0.8]{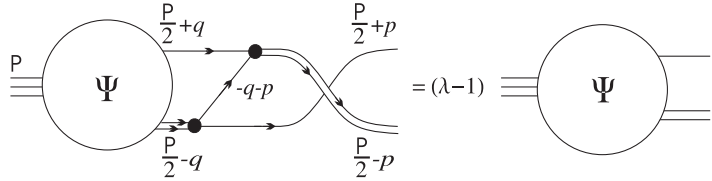}
\vspace{3mm}\caption{\small{The core baryon eigenvector  equation. The diquark vertex is
shown shaded. The diquark is  essentially the same size as the
baryon.}\label{figure:Nucleon}}
\end{figure}

The weightings are such that all the double helix diagrams may be summed 
to $TrLn({\bf 1}+K)-TrK=W_B-TrK$. 
Thus $Z[0,0]=\exp(W_B+D_R)$,
in which $D_R$ is the sum of the remaining connected diagrams.
To determine the content of $W_B$  we consider the
eigenvalue problem   $({\bf 1}+K)\Psi=\lambda\Psi$, 
which, for  ${\cal B}={\cal B}_{CQ}$,   has the following momentum space form, and is  illustrated
in Fig.\ref{figure:Nucleon}
\begin{equation}                                                             \label{eqn;A25}   
 \int \frac{d^4q}{(2\pi)^4}K(p,q;P)_{\alpha f, \gamma l}
^{\beta j, \rho h}\Psi_{\rho h}^{\gamma l}
(q;P)=(\lambda(P^2)-1)\Psi_{\alpha f}^{\beta j}(p;P), 
\end{equation}
\begin{equation}                                                              \label{eqn;A26}   
K(p,q;P)_{\alpha f,\gamma l}^{\beta j, \rho h}
= \sum_{i\delta }\frac{1}{12}
\Gamma_0\epsilon_{\gamma\alpha\delta}
 \epsilon_{lfi}i\gamma_5 C^TG^TC^Ti\gamma_5
\epsilon_{\beta\delta\rho} \epsilon_{jih}\Gamma_0 G\lambda^{-1}_0. 
\end{equation}

Equation (\ref{eqn;A25})  is a bound state equation for a three-quark  state in which the
paired quarks form a scalar diquark.  $\Psi$  in (\ref{eqn;A25}) is a Dirac spinor
and, as well, a  rank-2 tensor in both colour and flavour.  It should also be clear why the
baryon loop functional consists of twisted quark-diquark lines;  (\ref{eqn;A25}) is precisely of the form of
the Faddeev equations of standard three-particle theory - the twisting merely arose so that the
diagrammatic representation of (\ref{eqn;A25}) in Fig.\ref{figure:Nucleon} has  the same legs in the same
positions on both sides of the equation;   conventionally Fig.\ref{figure:Nucleon}  would  be drawn in
untwisted form. In fact (\ref{eqn;A25}) was also derived by conventional non-FIC three-particle
methods \cite{C89a}. See \cite{Lich,Ida} for early  quark-diquark models. We note that because (\ref{eqn;A20})  is
a discrete sum,  (\ref{eqn;A25}) is of separable form.
We separate   colour and flavour multiplets  in  (\ref{eqn;A25}), by decomposing
 $\Psi^{\gamma l}_{\rho h}$ 
according to
$({\bf 3} \otimes \overline{{\bf 3}})_C \otimes ({\bf 3} \otimes
\overline{{\bf 3}})_F= ({\bf 1}_C \otimes{\bf 1}_F) \oplus ({\bf 1}_C   \otimes {\bf 8}_F)\oplus({\bf
8}_C\otimes{\bf 1}_F)\oplus({\bf 8}_C\otimes{\bf 8}_F)$,
\begin{eqnarray*}
\Psi^{\gamma l}_{\rho
h}=\frac{1}{9}\delta_{\gamma\rho}[\Psi^{\alpha k}_{\alpha
k}]\delta_{lh}+\frac{1}{3}\delta_{\gamma \rho}[\Psi^{\alpha l}_{\alpha
h}-\frac{1}{3}\Psi^{\alpha k}_{\alpha k}\delta_{lh}]+
\end{eqnarray*}
\begin{equation}
   \mbox{\ \ \ \ \ \ \ \ \ \ + 
}\frac{1}{3}[\Psi^{\gamma k}_{\rho k}-\frac{1}{3}\Psi^{\alpha k}_{\alpha
k}\delta_{\gamma \rho}]\delta_{lh}+[\Psi^{\gamma l}_{\rho h}-\frac{1}{3}\Psi^{\alpha
l}_{\alpha h}\delta_{\rho \gamma}-\frac{1}{3}\Psi^{\gamma k}_{\rho
k}\delta_{lh}+\frac{1}{9}\Psi^{\alpha k}_{\alpha k}\delta_{\gamma \rho}\delta_{lh}].
\end{equation}  
\noindent Each  member($\Psi \equiv [....]$) of one multiplet is then seen to be an
eigenvector of
\begin{equation}                                                              \label{eqn;A27}   
\int \frac{d^4q}{(2\pi)^4}K(p,q;P)\Psi(q;P)
=(\lambda(P^2)-1) \Psi(p;P),\mbox{\ \ \ \ \ \ \ \ \ \ }
 \end{equation}
\begin{eqnarray}                                                                \label{eqn;A28}   
K(p,q;P)=-\frac{N[m]}{6}\Gamma_0(\frac{P}{4}+q+\frac{p}{2};\frac{P}{2}-p)
\Gamma_0(\frac{P}{4}+p+\frac{q}{2};\frac{P}{2}-q).  \nonumber \\
\mbox{\ \ \ \ \ \ \ \ \ \ \ \ \ \ \  \ \ \ \ \ \ \ \ \ \ \ \ }.\lambda_0 ((\frac{P}{2}-q)^2)^{-1}
G(-q-p)G(\frac{P}{2}+q),
\end{eqnarray}
where the $N[m]$ depend on the multiplet,
and we find $N[ {\bf1}_C\otimes{\bf 1}_F]= -2 ,\mbox{ } N[{\bf 1}_C\otimes{\bf 8}_F ]=+1$,  
$\mbox{ }N[{\bf  8}_C\otimes{\bf 1}_F]=+1$ and $N[{\bf 8}_C\otimes{\bf 8}_F ]= -\frac{1}{2}
.$  The ${\bf 1}_C\otimes {\bf 1}_F$ and ${\bf 8}_C\otimes {\bf 8}_F$
multiplets have negative values for $N$ and thus the quark rearrangement process is repulsive,
and the corresponding $\lambda(-M^2)$s have no zeros.  However the
colour octet - flavour singlet  `baryons'  have an $N$ value which means they are degenerate in
mass with the colour-singlet  flavour-octet baryons. Like the diquark states we expect colour non-singlets 
to not have a mass-shell, i.e. to be confined.  It is known that crossed gluon processes can perform this
task for the diquarks\cite{BRvS} (see sect.\ref{section:ConstituentMesonsandDiquarks}), but there has been no
corresponding analysis for these coloured baryonic states.

Let us now construct, for the colour singlet states,   an appropriate FIC representation  
for $\exp(TrLn({\bf 1}+K))$.
To this end note that  
an eigenvalue for positive energy solutions,
with degeneracy  2 (for spin $\uparrow$ and $\downarrow$), has the
form $\lambda_+^{\uparrow \downarrow}(P^2)=(M(P^2)+i\surd P^2a(P^2))F$,  
(define $F$ so that $a=1$ when $\lambda=0$, then $M_k(P^2)$ are baryon running masses),
while for negative energy solutions (anti-baryons)
$\lambda_-^{\uparrow \downarrow}=(\lambda_+^{\uparrow \downarrow})^{\star}$.
Thus, from the spectral representation for ${\bf 1}+K$,
\begin{equation}                                                            \label{eqn;A30}   
\exp(TrLn({\bf 1}+K))=\exp\left(\sum_k\hspace{-0.5em}n\int\hspace{-0.5em}d^4x\hspace{-0.5em}\int
\hspace{-0.5em}\frac{d^4P}{(2\pi)^4}[ln( \lambda_+^{\uparrow\downarrow}(P^2)^2)+
 ln(\lambda_-^{\uparrow \downarrow}(P^2)^2)]\right)
\end{equation}
where k sums the ground state and excited baryons states of (\ref{eqn;A25}), 
the squares in the $ln$ terms arise  from the spin degeneracy and $n=8$  arises 
from the  flavour degeneracy (the other baryon states are not shown).  
Therefore, with $a=1$ for simplicity,
\begin{eqnarray}                                                              \label{eqn;A31}   
\exp(TrLn({\bf 1}+K))=\exp\left(\sum_kn\int d^4x\int\frac{d^4P}{(2\pi)^4}
ln{\biggr [}(P^2+M_k(P^2)^2 )^2F_k^4{\biggr ]}\right) \nonumber \\ 
=\exp\left(\sum_kTrLn{\biggr [}(\gamma.\partial
+M_k({-\partial^2}))F_k^2\delta^4(x-y){\biggr ]} \right)  \nonumber \\ 
=\int D\overline{N}_kDN_k\exp\left(-\sum_k\int
d^4x\overline{N}_k(x) (\gamma.\partial+M_k({-\partial^2}))F_k^2N_k(x)\right),
\end{eqnarray}
in terms of $\overline{N}_k$ and $N_k$,  each of which is a flavour octet of
local baryonic spin $\frac{1}{2}$ FIC variables. Hence the exponentiated
sum of the closed double helix
diagrams is representable as a (free) baryon field theory.  The $F_k$ may be
absorbed with a re-definition of the  baryon fields.
Other more complicated (including baryon multi-loops) diagrams
 are present and constitute
a wealth of dressings and interactions between these (bare) baryons.

\subsection{Mesons}

We now briefly indicate how the non-diquark part of (\ref{eqn;A16}), $S[{\cal B}]$, gives the meson sector. 
The complete $Z$ has the form
\begin{eqnarray}                                                                 \label{eqn;A32}   
Z=\int D{\cal B}\exp\Bigl(-S[{\cal B}]
-\sum_{diquarks}TrLn(\lambda_k({-\partial^2};[{\cal B}_{CQ}])\delta^4(x-y))  \nonumber \\ 
\mbox{\ \ \ \ \ \ \ \ \ \ \
\ \ \ \ \ \ \ \ \ \ }+\sum_{baryons}TrLn((\gamma.\partial+M_k({-\partial^2} ;[{\cal
B}_{CQ}]))\delta^4(x-y))+....\Bigr).
\end{eqnarray} 
 We first determine  the dominant  configuration (and  constituent quark effect) ${\cal B}_{CQ}$, as the 
solution of the  Euler-Lagrange equations $\frac{\delta [S+..]}{\delta {\cal B^{\theta}}}=0$, which gives
\begin{equation}                                                                         \label{eqn;A33}   
{\cal B}_{CQ}^{\theta}(x,y)=D(x-y)\left[tr(G(x,y,{\cal B}_{CQ}) \frac{M_m^{\theta}}{2})+....
\right],
\end{equation} 
a non-linear equation for the $\{{\cal B}_{CQ}^{\theta}\}$. - a
Dyson-Schwinger type equation, where `+....'  are the diquark and baryon parts.  Dynamically it
describes the extensive self-energy of the quarks due to dressing by gluons, leading to the quark
running mass. This is finite
and extends over distances comparable to hadronic sizes and with an energy density that implies
that it has the dominant role in determining hadron structure. Only colour-singlet
translation-invariant solutions (depending only on $x-y$) are known.  Fourier and inverse
Fierz transforming (\ref{eqn;A33}) we obtain, on retaining only the meson contributions, 
\begin{equation}                                                                        \label{eqn;A34}   
 {\cal  B}_{CQ}(p)=c.\frac{4}{3}\int \frac{d^4q}{(2\pi)^4}D(p-q)\gamma_{\mu}\frac{1}{i\gamma.q+{\cal
M}+{\cal  B}_{CQ}(q)}\gamma_{\mu}.
\end{equation} 
Here $c=\frac{3}{4}$ from the
meson-diquark bosonisation, while normal Feynman rules would give $c=1$, as does the ${\bf 1-8}$ GCM
bosonisation \cite{CR85}. That $c < 1$  in the ${\bf 1-\overline{3}-3}$ bosonisation indicates that some strength
is generated by additional mechanisms in (\ref{eqn;A33}) involving meson and diquark processes, i.e. the
`book-keeping'  is more subtle in this bosonisation, and is so to avoid double counting. At present we
adopt the practice of using $c=1$ until these additional processes can be included dynamically.
 This CQ equation has unique  solutions when ${\cal M}\neq 0$
and 
  $G$ has the form $G(q)=[iA(q)q.\gamma+{\cal M}+B(q)]^{-1}$.
Expanding $S[{\cal B}]$ about its minimum gives $S[{\cal B}]=\sum_{n=0,2,3..}
S_n[{\cal B}]$, \ where $S_n$ is of order $n$ in ${\cal B}$ and, for example,  
$S_2= \frac{1}{2}\int {\cal B}^{\theta}(\Delta^{-1}_m)^{\theta \psi}{\cal B}^{\psi}$.
Introducing  bilocal source terms in (\ref{eqn;A32}) we have,
with $S^{\prime}=S-S_2$, and showing only the meson part,
\begin{eqnarray*}                                                             
Z[J]&=&\int D{\cal B}\exp(-S^{\prime}[{\cal B}]-S_2[{\cal B}]
        +\int {\cal B}^{\theta}J^{\theta}) \\ 
     &=&\exp(-S^{\prime}[\frac{\delta }{\delta J}])\int D{\cal B} \exp(-
       \int \frac{1}{2} {\cal B}^{\theta}(\Delta^{-1}_m)^{\theta \psi}
{\cal B}^{\psi}+ \int {\cal B}^{\theta}J^{\theta}), \\ 
&=&\exp(-S^{\prime}[\frac{\delta}{\delta j}])\int  Dm_k
       \exp(-\sum_k\frac{1}{2}\int m_k(x)
\lambda_k({-\partial^2})m_k(x)    +\int  j_km_k).
\end{eqnarray*}
Here we have used techniques similar to that for the diquarks and $\{m_k(x)\}$ is an infinite set of
local meson fields. Each ${m_k}$ corresponds  to one physical meson type, and the $\lambda_k$ are the 
eigenvalues  of the meson form of (\ref{eqn;A19})- a Bethe-Salpeter equation, which also gives the meson
form factors $\Gamma_k(p,P)$.
Applying the functional operator $\exp(-S^{\prime}[\frac{\delta}{\delta j}])$,
\begin{equation}                                                                        \label{eqn;A36}   
Z=\int Dm_k\exp\left(-\sum_k\frac{1}{2}\int
m_k(x)\lambda({-\partial^2})m_k(x)- S^{\prime}[m_k]\right) . 
\end{equation}
By  evaluation of $S^{\prime}[m_k]$, and identifying the mesons by their
quantum numbers, we obtain the  full local FIC representation for the meson sector of QCD
\cite{C89a,PRC87a,PRC87b}, as summarised in sect. \ref{subsection:HadronicLaws}.

\subsection{Hidden Chiral Symmetry\label{subsection:HiddenChiralSymmetry}}

When the quark current masses ${\cal M}
\rightarrow 0$ the fundamental action  $S[\overline{q},q,A^a_{\mu}]$ has the important
additional global $U_L(N_F)\otimes U_R(N_F)$ chiral symmetry.  The consequences of this follow naturally
through the GCM hadronisation.   The first significant result is that the dominant configuration
(\ref{eqn;A34}) has degenerate solutions
\cite{CR85,C89a,RCP88}
\begin{equation}                                                                         \label{eqn;A37}   
 G(q;V)=[iA(q)q.\gamma+VB(q)]^{-1}=\zeta^\dagger G(q;{\bf 1})\zeta^\dagger,
\end{equation}
where  $\zeta=\surd V, \mbox{\ \  }
V=\exp(i\surd2\gamma_5\pi^aF^a)$ and $\{\pi^a\}$ are arbitrary real constants
$\mid\!{\bf \pi}\!\mid \in [0,2\pi]$.  Thus
in the chiral limit the dominant configuration is degenerate and is the
manifold $(U_L\otimes U_R)/H$ (a coset space) where 
$H=U_V\subset U_L\otimes U_R$. Thus the chiral symmetry is represented as a hidden
symmetry.  This occurs because the action in (\ref{eqn;A32}) has the form of a `Mexican-hat' in the relevant
variables \cite{C89a}. The Nambu-Goldstone (NG) bosons form homogeneous Riemann coordinates for this
dominant configuration manifold. 
There is a technical complexity in expanding $S'[m_k]$  in the chiral limit which is caused by the action
having  degenerate minima, since  we do not have a unique minimum about which to expand.
First we need fields adapted to the compact dominant configuration manifold. For this we use the 
angles
$\{\pi\}$ as new field variables $\{\pi(x)\}$ \cite{RCP88,Pi94} in
place of some of the ${\cal B}^{\theta}(x,y)$. The dependence of the action on these dominant configuration
variables is such that the action will  increase only if the dominant configuration point is different at
different
 space-time points, and so a derivative expansion in
$\partial_{\mu}V(x)$ must arise.  The Dirac algebra  allows finally the
use of the matrix $U(x)=\exp(i\surd 2\pi^a(x)F^a)$ where
$V(x)=P_LU(x)^{\dagger}+P_RU(x) =\exp(i\surd 2\gamma_5\pi^a(x)F^a)$.  Then the NG
sector of $S[{\cal B}]$  is
\begin{eqnarray}                                                                  \label{eqn;A40}   
&& \int d^4x( \frac{f_{\pi}^2}{4}tr(\partial_{\mu}U\partial_{\mu}U^
{\dagger})+\kappa_1tr(\partial^2U\partial^2U^{\dagger})+ 
\kappa_2tr([\partial_{\mu}U\partial_{\mu}U^{\dagger}]^2) \nonumber \\ 
&& \mbox{\ \ \ \ \ \ \ \ \ }+\kappa_3tr(\partial_{\mu}U\partial_{\nu}U^{\dagger}\partial_{\mu}U
\partial_{\nu}U^{\dagger})+
\frac{\rho}{2}tr([{\bf 1} -\frac{U+U^{\dagger}}{2}]{\cal M})
+....
\end{eqnarray}
where $f_{\pi}, \kappa_1,..$ are given by explicit integrals \cite{RCP88,Pi94} in terms of $A(q)$ and
$B(q)$ and $\rho$ is the quark condensate parameter. In the chiral limit it is important 
to note \cite{C89a,RCP88}
that $B(q)$ in  the quark correlator is also the NG boson on-mass-shell form factor,
$\Gamma_{\pi}(p;{\bf P}={\bf 0})=B(p)$. In the chiral limit the ground state pseudoscalars 
play a dual role: they are at the same time both the NG bosons associated with the hidden
chiral symmetry  and also $\overline{q}q$ bound
states.  Under a chiral transformation  we find \cite{RCP88} $U(x)\rightarrow
U_LU(x)U_R^{\dagger}$.  This is a {\it derived} result of the FIC analysis which is usually
{\it assumed} in phenomenological modelling. We have included the lowest order term which depends on
${\cal M}$, i.e. for small breaking of the chiral symmetry by the quark current masses.  

The coupling of the baryon states to the above mesons requires us to keep the full ${\cal
B}$ in analysing the baryon sector, and not just the dominant configuration value ${\cal B}_{CQ}$. However
the  long wavelength limit of the NG-boson-baryon coupling may be inferred from the chiral
invariance of (\ref{eqn;A32}). Now $$TrLn{\biggr [}(\gamma.\partial +M({-\partial^2}))\delta^4(x-y){\biggr ]}=
TrLn{\biggr [}(\gamma.\partial+{\cal V}M({-\partial^2}))\delta^4(x-y){\biggr ]}$$ reflects that
invariance in (\ref{eqn;A32}), where ${\cal V}=\exp(i\surd 2\gamma_5 \pi^aT^a)$, with $\{T^a\}$ the
generators of $SU(3_f)$ ${\bf 8}$ representation.

\subsection{Hadronic Laws\label{subsection:HadronicLaws}}

Gathering the above results and   keeping only the low
orders in the hadronic variables and in the  derivatives (appropriate to
a low-energy long-wavelength expansion) 
\begin{eqnarray}                                                 \label{eqn;A41}           
Z=\int D\pi D\rho D\omega ..D{\overline N}DN...
\exp\left(-S_{had}[\pi,\rho,\omega,..\overline{N},N..]\right),
\end{eqnarray}                                                          
\begin{eqnarray}    \label{eqn;A42}                                                                          
S_{had}\hspace{-1em}& &\hspace{-1.5em}[\pi, \rho, \omega,...,\overline{N},N,..] =\hspace{5em}
\nonumber
\\  
 & &\int d^4x tr\{\overline{N}(\gamma.\partial+M_0+  \Delta
M_0-  M_0\surd 2i\gamma_5\pi^a{\cal T}^a+..)N\}\nonumber
\\                                                      
\mbox{\ \ \ \ \ \  \ \ \ \ }&+&\int d^4x\biggr[ \frac{f_{\pi}^2}{2}
[(\partial_{\mu}\pi)^2+m_{\pi}^2\pi^2] +
\frac{f_{\rho}^2}{2}[ -\rho_{\mu}(-\partial^2)\rho_{\mu} +(\partial_{\mu}\rho_{\mu})^2 
+m_{\rho}^2\rho_{\mu}^2] \nonumber \\ 
&+&\frac{f_{\omega}^2}{2}[\rho \rightarrow \omega]-f_{\rho}f_{\pi}^2g_ {\rho\pi\pi}\rho_{\mu}
.\pi \times\partial _{\mu}\pi-if_{\omega}f_{\pi}^3\epsilon_{\mu\nu\sigma\tau}\omega_{\mu}\partial_{\nu}
 \pi . \partial_{\sigma}\pi\times \partial_{\tau}\pi \nonumber \\ 
&-&if_{\omega}f_{\rho}f_{\pi}
G_{\omega\rho\pi}\epsilon_{\mu\nu\sigma\tau}\omega_{\mu}\partial_{\nu}
\rho_{\sigma}.\partial_{\tau}\pi \nonumber \\ 
&+&\frac{\lambda i}{80\pi^2}\epsilon_{\mu\nu\sigma\tau}tr(\pi.F\partial_{\mu}
\pi.F\partial_{\nu}\pi.F\partial_{\sigma}\pi.F\partial_{\tau}\pi. F)
+......\biggr],
\end{eqnarray}                                                        
in which the baryon octet is finally written as a  rank-2 tensor,
$N=N^a{\cal T}^a$, where the $\{{\cal T}^a\}$ are
  generators of the $SU(3_f)$ ${\bf 3}$ representation. 
We have written $\lambda_j(P^2)=(P^2+m_j(P^2)^2)f_j^2$
 where $m_j(P^2)$ are the running meson masses,  but  only the physical
masses  (from $\lambda(P^2)=0$) are shown above. The imaginary terms in this
meson action are the chiral anomalies of QCD, including in particular the Wess-Zumino term.
In (\ref{eqn;A42}) we have shown $m_{\pi}$ and $\Delta M_0$ which are mass terms from the
 chiral symmetry breaking quark current masses, while $M_0$ is the
`chiral mass' of the constituent baryons (see sect.\ref{section:ConstituentNucleon}). For non-zero
quark current masses the NG boson masses $\{m_{\pi}\}$ and the  baryon octet mass splittings $\{\Delta
m_0\}$ are seen to satisfy the  Gell-Mann-Okubo and Coleman-Glashow formulae \cite{CRP89}.

In general the coupling terms in the hadronic action are non-local and  the actions in
(\ref{eqn;A40}) and (\ref{eqn;A42}) will also contain higher order derivative terms like 
$tr\{\overline{N}(
m_N\surd 2i\gamma_5\partial^2\pi^a{\cal T}^a+..)N\}$. These should not be thought of as `different'
couplings, but rather as just arising from the expansion of the meson-baryon vertex function
$\Gamma_0(p,q)$.   Hence
rather than making this effective action non-renormalisable, as often occurs  in effective actions, 
such terms when properly retained as parts of  complete vertex functions actually render 
loop diagrams finite (an example is the pion-nucleon loop
in sect.\ref{section:PionDressingoftheNucleon}).

The full non-local meson sector of (\ref{eqn;A42}) \cite{PRC87b} has been used in many studies, such as
the
$\rho\rightarrow\pi\pi$ decay \cite{PRC87a},
$\omega-\rho$ mass splitting  \cite{RCP89}, 
charge symmetry breaking via $\rho-\omega$ mixing \cite{Mitchell:1994jj},
pion loop contribution to $\rho-\omega$ mixing  \cite{Mitchell:1996dn} and 
pion and $\rho$ meson observables \cite{Frank},  for extensions to include electromagnetic
interactions  \cite{Frank:1993ye,Hecht:1997uj}, and for $\eta$ and $\eta'$ \cite{KK97}. For the pion loop
contribution to the electromagnetic pion charge radius see \cite{Alkofer:1995gu}.  The chirally invariant
form of the NG boson sector in (\ref{eqn;A40}) in particular has been investigated in  
\cite{Frank:1996yb} and  the $\pi-\pi$ scattering lengths in \cite{Pi94}.

\section{Effective Gluon Correlator\label{section:EffectiveGluonCorrelator}} 

We now consider the  detailed implementation of the GCM.
First we must determine the effective gluon correlator  by fitting GCM observables to experimental data.
This involves the determination of the dominant configuration, i.e. the constituent quark effect. This
effective gluon correlator  and quark correlator  are then used in the constituent-meson BSE equations
in order to determine meson masses, and also $f_{\pi}$. 
The dominant configuration  is defined by equations,
\begin{equation}                                                                    \label{eqn;R8}
\frac{\delta S}{\delta {\cal B}(x,y)}\left|_{{\cal B}_{CQ}}=0\right.. 
\end{equation}
Of the set ${\cal B}_{CQ}(x,y)$ only  $A(x-y)$ and $B(x-y)$ (their Fourier transforms appear in
(\ref{eqn;R10})) are non-zero bilocal fields characterising the
dominant configuration. They are also  translation-invariant.  This is the dynamical breaking
of chiral symmetry. Such non-zero dominant configurations are also known as condensates. 
 Writing out the translation invariant CQ equations
 we find that the dominant configuration is indeed simply the constituent  quark
effect as they may be written  in the form of (\ref{eqn;A33}) or (\ref{eqn;A34}),  or
\begin{equation}                                                                  \label{eqn;R9}
G^{-1}(p)=i\backslash \!\!\!p +m+
\frac{4}{3}\int\frac{d^4q}{(2\pi)^4}D_{\mu\nu}(p-q)\gamma_{\mu}G(q)\gamma_{\nu},
\end{equation}
which is  the gluon dressing of a constituent quark.  Its solution has the structure  
\begin{equation}                                                                   \label{eqn;R10}
G(q)=(iA(q^2;m)q.\gamma+B(q^2;m)+m)^{-1}=-iq.\gamma\sigma_v(q^2;m)+\sigma_s(q^2;m).
\end{equation}
 In the chiral limit there are more ${\cal
B}_{CQ}$  fields that are non-zero, and the resultant degeneracy of the dominant
configuration is responsible for the masslessness of the pion.  
The  constituent quark  correlator  $G$ should not be confused with the 
 complete quark correlator ${\cal G}$ from (\ref{eqn;R1}) which would be needed
to analyse the existence or otherwise of free quarks. The $G$ on the other hand relates
exclusively to the internal structure of hadrons, and to the fact that this structure appears to be
dominated by the constituent quark effect. The evaluation of  ${\cal G}$ is a very difficult
task, even within the GCM, while $G$ is  reasonably easy to study using (\ref{eqn;R9}). The 
truncation of the DSE in which the full quark ${\cal G}$ is approximated by this $G$ amounts to
using a mean field approximation (see (\ref{eqn;B14})); however from the tDSE there is no
systematic formalism  for going beyond the mean field as there is in the GCM.
 The hadronic effective action in (\ref{eqn;A32}) arises when $S[{\cal B},..]$ is expanded
about the dominant CQ configuration; the first derivative is zero  by (\ref{eqn;R8}), and the
second derivatives, or curvatures, give the constituent or 
core meson correlators $G_m(q,p;P)$
\begin{equation}                                                                    \label{eqn;R11}
 G_m^{-1}(q,p;P)=F.T.\left(\frac{\delta^2 S}{\delta {\cal B}(x,y)\delta {\cal
B}(u,v)}\left|_{{\cal B}_{CQ}}\right.\right),
\end{equation}
after exploiting  translation invariance and Fourier transforming. Higher order
derivatives  lead to couplings between the  meson cores.  The $G_m(q,p;P)$ are given by 
ladder-type  correlator equations and the non-ladder effects are inserted by
the final functional integrals in  (\ref{eqn;R7}), giving the complete GCM meson correlators
${\cal G}_m(q,p;P)$.  It is interesting
to note that the truncated and modified DSE in Maris and Roberts\cite{MR} are identical to
 (\ref{eqn;R9}) and (\ref{eqn;R11}), in the form of (\ref{eqn;R12}). 

 In the fitting to meson data the $\omega$ and a$_1$ mesons are described by
these constituent meson correlators; that is, we ignore meson dressings of these mesons. 
The mass $M$ of these states is determined by finding the pole position of $G_m(q,p;P)$ in the
meson momentum $P^2=-M^2$ and this, or equivalently the meson version of (\ref{eqn;A19}) which for
the mass-shell has $\lambda(P^2)=0$, leads to the homogeneous vertex equation
\begin{equation}                                                                  \label{eqn;R12}
\Gamma(p;P)=-\frac{4}{3}\int\frac{d^4q}{(2\pi)^4}D_{\mu\nu}(q-p)
\gamma_{\mu}G(q+\frac{P}{2})\Gamma(q;P)G(q-\frac{P}{2})\gamma_{\nu}.
\end{equation} 

 To solve (\ref{eqn;R9}) for various $D_{\mu\nu}(p)$ and then to proceed to use 
$A(s)$ and $B(s)$ in meson correlator equations for fitting observables to meson data is
particularly difficult. A  robust numerical technique is to use a separable
expansion\cite{CG95a} as follows. We have in the Landau gauge
\begin{equation}                                                                 \label{eqn;R13}
D_{\mu\nu}(p)=(\delta_{\mu\nu}-\frac{p_{\mu}p_{\nu}}{p^2})D(p^2), \mbox{\ \ and \ \ } 
{\cal G}_{\mu\nu}(p)=(\delta_{\mu\nu}-\frac{p_{\mu}p_{\nu}}{p^2}){\cal G}(p^2),
\end{equation}
where $D(p^2)=g(p^2){\cal G}(p^2)g(p^2)$.
First expand   $D(p-q)$ in (\ref{eqn;R9}) into $O(4)$ hyperspherical harmonics
\begin{equation}                                                                   \label{eqn;R14}
D(p-q)=D_0(p^2,q^2)+q.pD_1(p^2,q^2)+...,
\end{equation}
\begin{equation}                                                                   \label{eqn;R15}
D_0(p^2,q^2)=\frac{2}{\pi}\int_0^{\pi}d\beta \mbox{sin}^2
 \beta D(p^2+q^2-2pq\mbox{cos}
\beta),...
\end{equation}
Introduce a multi-rank separable $D_0$ expansion (here $n=3$)
\begin{equation}                                                            \label{eqn;R16}
D_0(p^2,q^2)=\sum_{i=1,n} \Gamma_i(p^2)\Gamma_i(q^2),
\end{equation}
and the constituent quark equations then have solutions of the form (in the chiral limit)
\begin{equation}                                                             \label{eqn;R17}
B(s)=\sum_{i=1,n} B_i(s), \mbox{\ \  } B_i(s)=b_i\Gamma_i(s), 
\mbox{\ \  }\sigma_s(s)=\sum_{i=1,n}\sigma_s(s)_i, 
\end{equation}
\begin{equation}                                                               \label{eqn;R18}
 b_i^2=4\pi^2\int_0^{\infty} sds B_i(s)\sigma_s(s),
\mbox{\ \ \  and \ \ }
B_i(s)=\frac{\sigma_s(s)_i}{s\sigma_v(s)^2+\sigma_s(s)^2}.
\end{equation}

\begin{figure}[ht] 
\hspace{15mm}\includegraphics[scale=0.5]{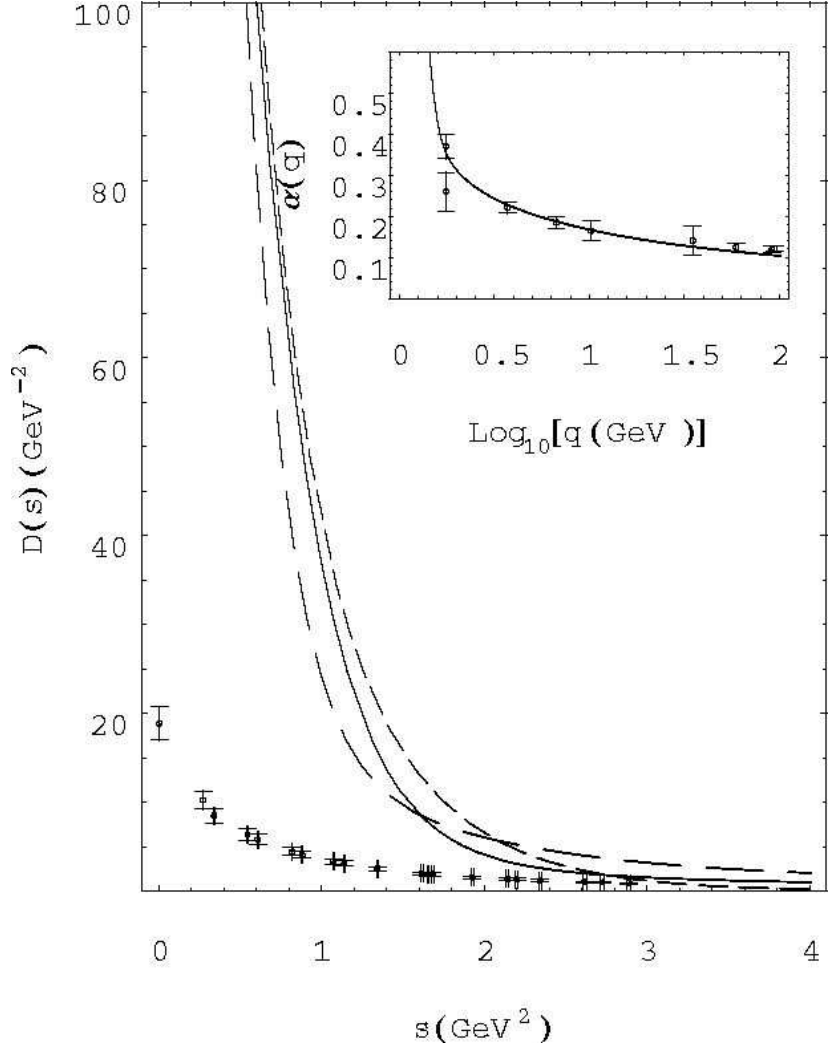}
\vspace{-3mm}\caption{\small{Plots of $D(s)$ (GeV$^{-2}$). Solid line is GCM98; shortdash line
is GCM95; longdash line is GCM97. Data plot is lattice pure-gluon ${\cal G}(s)$ from Marenzoni {\it et
al.}, and so has no quark-gluon vertex. Insert is fit of GCM98 to $\alpha(s)$ of Particle Data Book.}
\label{figure:BigPlot}} 
\end{figure}

However rather than specifying $\Gamma_i$ in (\ref{eqn;R16}) we  proceed by
parametrising forms  for the $\sigma_{si}$ and $\sigma_v$; the  $\Gamma_i$
then follow from (\ref{eqn;R17}) and (\ref{eqn;R18}): 
\begin{eqnarray*}
\sigma_s(s)_1=c_1\mbox{exp}(-d_1s), \mbox{\ \ \ \ }
\sigma_s(s)_2=c_2.\left(\frac{2s^2-d_2(1-\exp(-2s^2/d_2))}{2s^4}\right)^2,
\end{eqnarray*}
\begin{eqnarray*}
\sigma_s(s)_3=
c_3\left(\frac{2f(s)-d_3(1-\exp(-2f(s)/d_3))}{2f(s)^2}\right)^2,\mbox{\ }
f(s)=s(ln(\tau +s/\Lambda^2))^{1/2}
\end{eqnarray*}

\begin{equation}                                                       
\label{eqn;R19} 
\sigma_v(s)= 
\frac{2s-\beta^2(1-\exp(-2s/\beta^2))}{2s^2}.
\end{equation}
 The three $\sigma_{si}$ terms mainly determine the IR, 
midrange and UV regions; the $\sigma_s(s)_3$ term describes the 
asymptotic form of $\sigma_s(s)\sim 1/s^2\ln(s/\Lambda^2)$ for $s\rightarrow \infty$ 
 and ensures the  form for $D(s)\sim 1/s\ln(s/\Lambda^2)$. The parameter $\tau=10$ ensures that $\sigma_3$ is
well behaved at small $s$, but otherwise has no effect on the fit. 
With these parametrised forms  we can numerically  relate, in a robust 
and stable manner, the   parameter set in Table 1 to the mass of the 
$a_1$(1230MeV) and $\omega$(783MeV)  mesons from (\ref{eqn;R12}), to  
$f_{\pi}$(93.3MeV)  and to  experimental $\alpha(s)$ points (see insert in Fig.\ref{figure:BigPlot})  from
the Particle Properties Data Booklet for
$s >3$GeV$^2$.

The translation invariant form for the
effective gluon correlator is  easily reconstructed by using 
$D(p^2)=D_0(p^2,0)$ in (\ref{eqn;R14}) and then from (\ref{eqn;R16}) 
\begin{equation}                                                               \label{eqn;R20}
D(p^2)=\sum_i\frac{1}{b_i^2}\frac{\sigma_s(0)_i}{\sigma_s(0)^2}
\frac{\sigma_s(p^2)_i}{p^2\sigma_v(p^2)^2+\sigma_s(p^2)^2}.
\end{equation}

\begin{figure}[ht] 
\hspace{25mm}\includegraphics[scale=0.59]{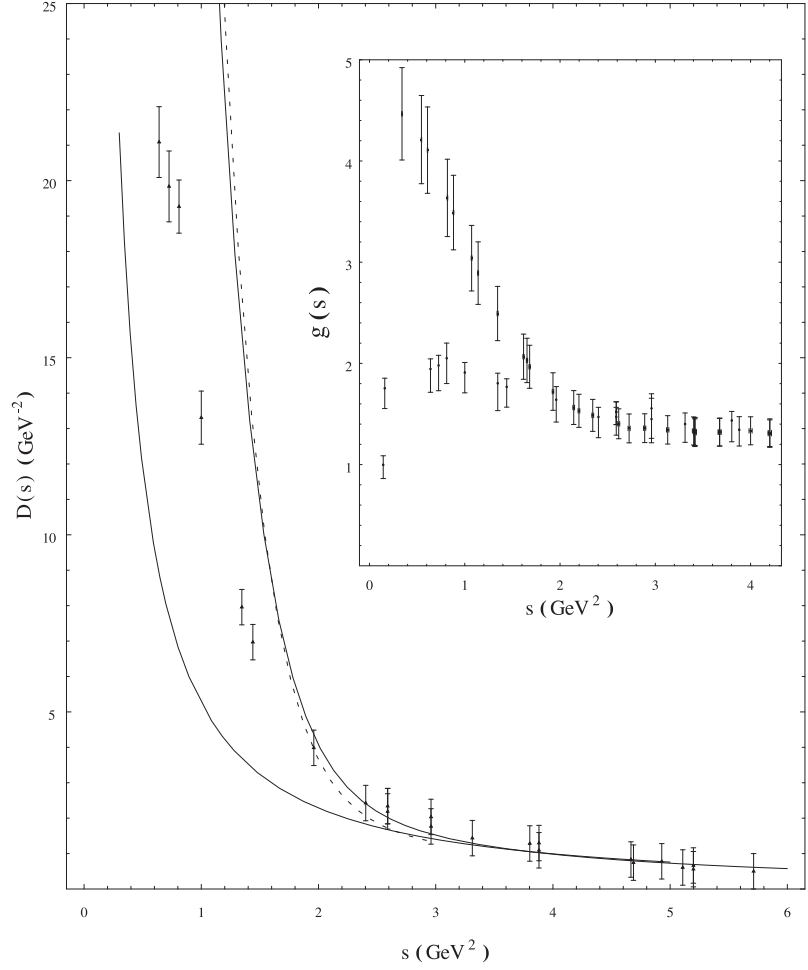}
\vspace{-3mm}\caption{\small{Plots of $D(s)$ (GeV$^{-2}$). GCM98 is the upper solid line which is almost 
identical to the Jain and Munczek $D(s)$ (dashed line); lower solid line is two-loop form with
$\Lambda=0.234$GeV, $N_f=3$ and  $\tau=10$; data plot is combined lattice data for
$g(s){\cal G}(s)g(s)$ with ${\cal G}(s)$ from Marenzoni {\it et al.} (as in (a)) and $g(s)$ from Skullerud.
Insert shows $g(s)$ from Skullerud (lower data plot), and from GCM98/${\cal G}(s)$ (upper data plot).}
\label{figure:SmallPlot}} 
\end{figure}

 With the parameter set in Table 1 the resulting quark-quark coupling correlator $D(p^2)$ is shown in
 Fig.\ref{figure:BigPlot} and  Fig.\ref{figure:SmallPlot}.  
 A significant feature of QCD is
that the infrared dominance, as revealed by the large value of $D(s)$ at small $s$, causes the
CQ equations to saturate, i.e. the forms of the solutions $A(s)$ and $B(s)$ at low $s$ are
independent of the detailed IR form of $D(s)$.  This saturation effect means  that low energy QCD is
surprisingly easy to model, and this effect is utilised in the GCM.

In Fig.\ref{figure:SmallPlot} we plot the new GCM98 quark-quark coupling correlator
$D(s)$  which shows excellent overall agreement with the Jain-Munczek $D(s)$\cite{JM,KK}, and we
also plot 
 $D(s)=g(s){\cal G}(s)g(s)$ constructed from the  Marenzoni {\it et al.} \cite{Marenzoni} ${\cal
G}(s)$ and Skullerud \cite{Skul98} $g(s)$ lattice results which shows agreement down to $s=1.8$GeV$^2$. 
The normalisation
and shape of the Marenzoni {\it et al.} ${\cal G}(s)$  agrees with that of Leinweber {\it et
al.}
\cite{LSWP}. However  the normalisation of the much more difficult lattice computation of  $g(s)$ 
is uncertain and we have chosen it  so that the combined lattice $D(s)$ agrees with the
experimental Particle  Properties Data Booklet for $s>3$GeV$^2$. As shown  in
Fig.\ref{figure:SmallPlot}  all three $D(s)$  depart from the two-loop form below  $s=2.5$GeV$^2.$ 
The difference between the GCM98 (or Jain-Munczek) and the lattice construction could be an
indication of contributions to the quark-quark coupling  from higher order
 gluon self-couplings at low energy, since processes like  Fig.\ref{figure:Ints}(c) would be
included in the GCM fit, but are not in the lattice construction. The earlier GCM95 and GCM97 gave $D(s)$
that differed mainly in the asymptotic region, see plots in \cite{CG98a}.  

The insert in Fig.\ref{figure:SmallPlot} shows the $g(s)$ from 
$g^2(s)=D(s)/{\cal G}(s)$  that then follows from our analysis. This is the
effective quark-gluon coupling vertex   if the gluon correlator is taken to be that of 
Marenzoni {\it et al.} (or Leinweber {\it et al.}) and here the error bars now indicate combined
errors and uncertainties from the lattice spacing. Also shown is
$g(s)$ from Skullerud\cite{Skul98} with the  normalisation as discussed above.  
 
We summarise in Table 2 some of the hadronic observables that may be computed  within the GCM,
showing in particular their sensitivity to the evolving modelling of the effective gluon correlator.
Further meson observables are reviewed in Tandy \cite{Tandy}.

.

\vspace{8mm}
\hspace{8mm} {\bf Table 1:  GCM1998 $\sigma_s(s)$ and $\sigma_v(s)$  Parameters.}

\vspace{2mm}
\hspace{4mm}
\begin{tabular}{|llllll|}
\hline 
$\mbox{ \ \ }$$c_1$ &1.839GeV$^{-1}$  & $d_1$    & 3.620GeV$^{-2}$&$\mbox{ \ \ }\beta$  & 0.4956GeV
\\
$\mbox{ \ \ }$$c_2$ &0.0281GeV$^{7}$  & $d_2$    & 1.516GeV$^{4}$&$\mbox{ \ \ }\Lambda$ &0.234GeV\\
$\mbox{ \ \ }$$c_3$  & 0.0565GeV$^{3}$  &$d_3$  & 0.7911GeV$^2$ & & \\  
\hline
\end{tabular}
\vspace{3mm}

\newpage
.
\newpage

 {\bf Table 2:  Hadronic Observables.}

\vspace{2mm}
{
\hspace{-3mm} \begin{tabular}{| l r r r r |} 
\hline 
 {\bf Observable} & {\bf GCM1995}& {\bf GCM1997}& {\bf GCM1998} & {\bf
Expt/Theory} \\
\hline  
$f_{\pi}$  &    93.0MeV*& 93.2MeV*&  92.40MeV*  &93.3MeV\\
$a_1$ meson mass  & 1230MeV* &1231MeV* & 1239MeV* &1230MeV\\
$\pi$ meson mass(for m$_{u,d}$)   &  138.5MeV*   & 138.5MeV* &  138.5MeV*  &138.5MeV\\
     $\alpha(s)$ &- & - &fitted & $\dagger$\\
$K $ meson mass (for $m_s$)   &  496MeV*      & -  &-&     496MeV\\
$(m_u+m_d)/2|_R(\mu=1$GeV)   &  6.5MeV & 4.8MeV  &7.7MeV &  $\approx$8.0MeV\\
$m_s|_R(\mu=1$GeV)   &  135MeV   & - &  - &130MeV\\
$\omega$ meson mass &  804MeV  & 783MeV* & 783MeV*  &782MeV  \\
$a^0_0$ $\pi-\pi$ scatt. length &  0.1634  & 0.1622 &  0.1657 &   0.26 $\pm$ 0.05 \\
$a^2_0$ $\pi-\pi$ scatt. length&  -0.0466  &-0.0463 & -0.0465  & -0.028 $\pm$ 0.012 \\
$a^1_1$  $\pi-\pi$ scatt. length&  0.0358   & 0.0355 & 0.0357  &0.038 $\pm$ 0.002 \\
$a^0_2$ $\pi-\pi$ scatt.length&  0.0017   & 0.0016& 0.0018  &0.0017 $\pm$ 0.003\\
$a^2_2$ $\pi-\pi$ scatt.length &  -0.0005  &-0.0005 & -0.0003  &.00013$\pm$0.0003\\
$r_{\pi}$ pion charge radius & 0.55fm & 0.53fm & 0.53fm  &0.66fm\\
$\frac{1}{2}^+(0^+)$nucleon-core mass$^{**}$  &  1390MeV & 1435MeV & 1450MeV  
&$>$1300MeV$\dagger\dagger$  \\
const. quark rms size   & 0.51fm&0.39fm & 0.58fm  &-\\
chiral quark const. mass &   270MeV& 267MeV  & 325MeV  &-\\
$0^+$ diquark rms size   & 0.55fm&0.55fm & 0.59fm  &-\\
$0^+$ diquark const. mass &  692MeV  & 698MeV & 673MeV  &$>$400MeV  \\
$1^+$ diquark const.  mass &  1022MeV  &903MeV & 933MeV  & -  \\ 
$0^-$ diquark const.  mass &  1079MeV  & 1049MeV & 1072MeV  &-  \\
$1^-$ diquark const.  mass &  1369MeV  & 1340MeV & 1373MeV  &-  \\
MIT bag constant  &  (154MeV)$^4$  &(145MeV)$^4$ & (166MeV)$^4$  &(146MeV)$^4$ \\
MIT N-core (no cms corr.)   & 1500MeV & 1420MeV  & 1625MeV  &$>$1300MeV$\dagger\dagger$   \\
\hline 
\multicolumn{5}{|l|}{* fitted observable;  -  not computed or not known; $\dagger$
$\alpha(s)$ from Particle Properties}\\ 
\multicolumn{5}{|l|}{Data Booklet; GCM1995: \cite{CG95a};  GCM1997:  
\cite{CG97b}; GCM1998: \cite{CG98a}.}\\
\multicolumn{5}{|l|}{$^{**}$ only $0^+$ diquark correlation; $1^+$ diquark correlation
lowers nucleon core mass.}\\
\multicolumn{5}{|l|}{ $\dagger\dagger$ nucleon core mass (i.e. no meson dressing). }\\
\hline
\end{tabular}}

\normalsize

\section{Constituent Quarks\label{section:ConstituentQuarks}}

 The constituent quark effect  \cite{PCR88} is
the dominant effect in determining the structure of hadrons, and also their response in
scattering events, particularly deep inelastic  scattering. The constituent mass effect 
manifested itself in the early studies of  baryon magnetic moments. 
 The GCM analysis reveals
rather directly both the effective mass and the effective size of the constituent quarks, and
 relates these  to the effective gluon correlator. We consider chiral limit
constituent quarks and we carefully define constituent quarks and constituent hadrons as those
constructs appearing in exponentiated effective actions in the functional formulation of the GCM
and distinguish them from the exact correlations, which follow from the complete functional
integrations.
We can express the full quark correlator ${\cal G}(x,y)$ in terms of the bosonised
FIC variables, with  (\ref{eqn;R1}) and (\ref{eqn;A16}) giving
\begin{equation}                                                            \label{eqn;B14}
{\cal G}(x,y)=\frac{\int D{\cal B}D{\cal D}D{\cal D}^{\star}G(x,y,{\cal B})
                  \mbox{exp}(-S[{\cal B},{\cal D},{\cal D}^{\star}])}
                {\int D{\cal B}D{\cal D}D{\cal D}^{\star}
                  \mbox{exp}(-S[{\cal B},{\cal D},{\cal D}^{\star}])},
\end{equation}
where  $G(x,y,{\cal B})$ is defined in   (\ref{eqn;A15}).           
The bilocal field functional integrals can be further decomposed into 
local hadronic functional integrals. 
The constituent quark effect appears as the dominant configuration about 
which the meson-diquark 
bosonised GCM action in (\ref{eqn;B14})  (and elsewhere) is expanded. This is characterised by 
${\cal D} = 0$ and  two of the ${\cal B}\neq 0$ (the ${\cal
B}_{CQ}$).  At the
dominant configuration
$G(x,y,{\cal B}_{CQ})$  is the constituent quark correlator which appears in the constituent  meson and
baryon  correlators.  The  ${\cal B}_{CQ}$ eqns.(\ref{eqn;A34}) have the form (we   use a
Feynman-like gauge to simplify the discussion, but the more realistic Landau gauge may  be used), 
\begin{equation}                                                                \label{eqn;B15} 
B(p^2)=\frac{16}{3}\int\frac{d^4q}{(2\pi)^4}D(p-q).\frac{B(q^2)}
{q^2A(q^2)^2+B(q^2)^2},
\end{equation}
\begin{equation}                                                                \label{eqn;B16}
[A(p^2)-1]p^2=\frac{8}{3}\int\frac{d^4q}{(2\pi)^4}q.pD(p-q).\frac{A(q^2)}
{q^2A(q^2)^2+B(q^2)^2}.
\end{equation}
 Here $B(q^2)$ and  $C(q^2)=A(q^2)-1$ are
the Fourier transforms of the  two 
${\cal B}_{CQ}(x,y)$  fields, and where translational invariance
is used: ${\cal B}_{CQ}(x,y) \rightarrow {\cal B}_{CQ}(x-y)$.
We see from
(\ref{eqn;B14}) that the full quark correlator is given by the dressing of the constituent quark
correlator by various hadronic fluctuations. In the context of the
 tDSE   (\ref{eqn;B15}) and (\ref{eqn;B16}) (in Landau gauge) are  often used as an approximation \cite{MR}
to the full quark correlator; ${\cal G}(x,y) \approx G(x,y,{\cal B}_{CQ})$, thereby confusing the full and
constituent quark correlators.   The GCM formulation clearly reveals that the tDSE approach is actually
using a mean field approximation. The additional  processes manifest in (\ref{eqn;B14}) show up
in the FIC hadronisation as the dressing of constituent hadrons by other hadrons, of which the pion dressing
of the nucleon is the most pronounced example. These dressings incorporate additional processes
corresponding to further dressing of the constituent quarks  as well as 
  further  interactions between the constituent quarks.

When considering only the constituent quarks the bilocal effective action in  
(\ref{eqn;B14})  may be simplified  by keeping  only the translation invariant $B$ and $C$
dependence. This leads to a reduced  action per quark flavour,
\begin{eqnarray*} 
S_{CQ}[B,C]=V\left(-\frac{12\pi^2}{(2\pi)^4}\int dq q^3
\mbox{ln}(A^2(q^2)q^2+B(q^2)^2)\right.
\end{eqnarray*}
\begin{equation}                                                                \label{eqn;B17}
\hspace{40mm}+\left.\frac{9}{4}\pi^2\int dx x^3
\frac{B(x)^2}{D(x)}+\frac{9}{2}\pi^2\int dx x^3
\frac{C(x)^2}{D(x)}\right)
\end{equation} 
where  $V$ is the
(infinite) spacetime volume.  The minimization of $S_{CQ}[B,C]$ gives
(\ref{eqn;B15}) and (\ref{eqn;B16}).   The action $S_{CQ}[B,C]$ has the form of a sum of a {\em kinetic 
energy} term  (defined as that part which is local in momentum) and a {\em potential energy} term (defined as
that part which is local in relative spacetime). Both have unconventional
forms because (\ref{eqn;B17}) describes self-interaction effects. The kinetic energy term
involves the constituent quark running mass $M(s)=B(s)/A(s)$. 

\begin{figure}[ht]
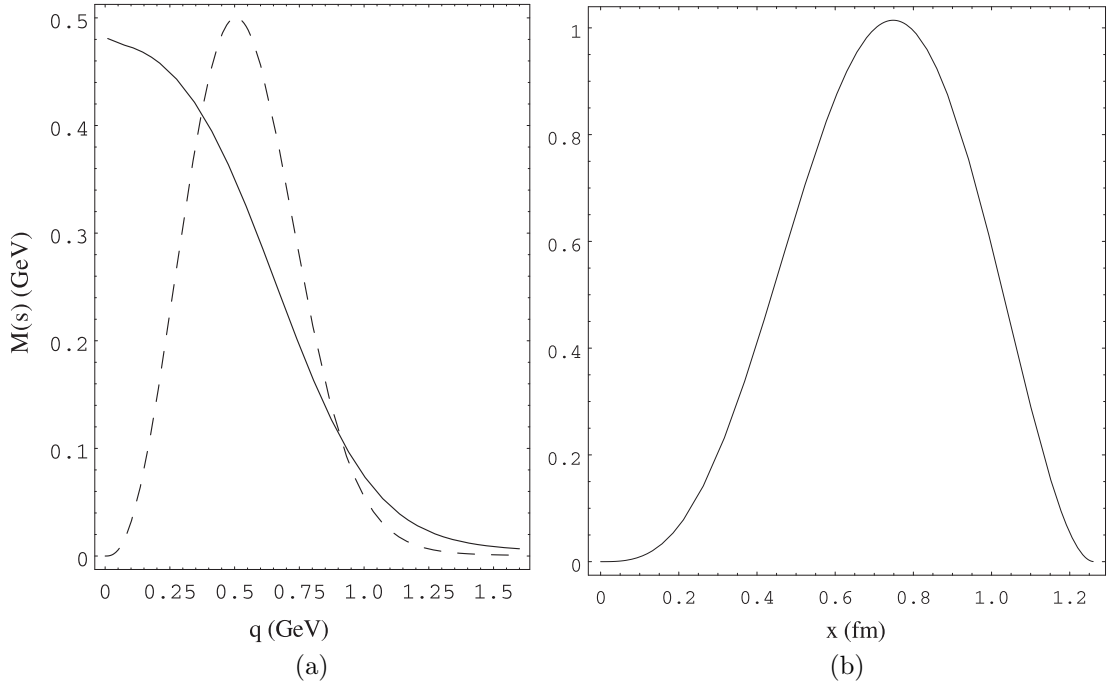

\vspace{-5mm}\caption{\small{ (a) Quark running mass $M(s)$ (solid line) and the integrand of the kinetic
energy part of the constituent quark action. (b) Integrand  of the potential energy part of the
constituent quark action.}
 \label{figure:KEPEaction}}
\end{figure}

  Since the key constituent quark effective mass is
associated with the kinetic part of (\ref{eqn;B17}), we subtract  the $B=0$ form. In this way we
compare the  non-perturbative configuration with the perturbative configuration, and it is this
difference which also generates the MIT bag constant discussed in
sect.\ref{section:ConnectiontoOtherModels},

\begin{eqnarray}                                                                \label{eqn;B18} 
S_{CQ}[B]=V\biggr(-\frac{12\pi^2}{(2\pi)^4} \int_0^{+\infty} dq \left[q^3
\mbox{ln}\left(\frac{A(q^2)^2q^2+B(q^2)^2}{ A^2(q^2)q^2}\right)\right] +\frac{9}{4}\pi^2\int_0^{+\infty} dx 
\left[x^3\frac{B(x)^2}{D(x)}\right]\biggr).
\end{eqnarray} 
The   kinetic energy and potential energy  integrands, indicated by the square brackets in
(\ref{eqn;B18}),  are shown in  Fig.\ref{figure:KEPEaction}. Fig.\ref{figure:KEPEaction}(a) also shows the
quark running mass.
 The chiral limit constituent quark mass of approximately  $300$MeV  is revealed
 as the value of the running mass that dominates the kinetic energy integrations.
The width of the $q$-integrations being a `fermi-motion' effect. The integrand  of the
potential energy term shows that gluon exchanges up to some 1.2fm are relevant.  Hence we see  directly that
the constituent quark characteristics  are implicit  in the
action  (\ref{eqn;B17}).  

\vspace{3mm}
\begin{minipage}[t]{50mm}

\hspace{-5mm}\includegraphics[scale=0.85]{cKEPlot.EPSF}

\makebox[70mm][c]{(a)}
\end{minipage}
\begin{minipage}[t]{50mm}
\hspace{15mm}\includegraphics[scale=0.765]{cPEPlot.EPSF}

\makebox[110mm][c]{(b)}
\end{minipage}

\section{Constituent Mesons and Diquarks\label{section:ConstituentMesonsandDiquarks}}

\begin{figure}[ht]
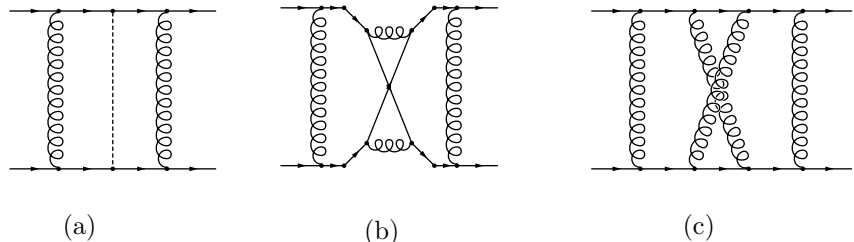

\vspace{-7mm}
\caption{\small{ (a) Diagram  shows a meson exchange from the functional integral in
(\ref{eqn;B17}), dressing constituent ladder diquark correlations.  (b) a low order gluon process
within the meson exchange. In (c) we redraw (b)  to show the crossed gluon
processes inserted via meson exchange.}
 \label{figure:crossed}}
\end{figure}

From the hadronisation we saw that the constituent mesons and diquarks arise as 
ladder  BSE states; with their correlators described as the generalised curvatures of the meson-diquark
boson action when expanding about the dominant or constituent quark configuration.  This provides
a  particularly instructive insight into the true role of such ladder BSE states.  Traditionally these
ladder states arose as a severe truncation of the full coupled DSE,  however we see that they actually
play a key dynamical role in the hadronisation in that they naturally arise as appropriate  FIC variables, rather
than from some unstructured approximation scheme. The crossed diagrams that are normally neglected in the
tDSE are automatically inserted in the GCM via the hadronic functional integrations in  (\ref{eqn;A41}),
and it is these  processes that convert the constituent mesons and baryons into the
observable hadronic modes, as seen later in sect.\ref{section:PionDressingoftheNucleon} for the
nucleon.  Hence the GCM hadronisation reveals a `book-keeping' that was not previously known.  In
this connection Bender, Roberts and von-Smeckal
\cite{BRvS} have found evidence that these additional crossed gluon processes may be responsible for
confining the diquarks, which is particularly interesting since the ladder BSE have a mass-shell for
the diquarks, and it is these masses which are shown in Table 2, see also 
\cite{Hellstern:1997nv}.   That study explicitly included the crossed diagrams by  extending the BSE equation to include the
lowest order crossed diagram.  However in the GCM hadronisation  such crossed diagrams are seen to arise from meson
exchanges, as shown in Fig.\ref{figure:crossed}.  Evidence from the nucleon computations  suggests that due to the
high `constituent mass' of the  diquarks the presence of the  pole in the diquark  correlator  at the
mass-shell is not dynamically significant.


\hspace{15mm}\begin{minipage}[t]{35mm}
\hspace{5mm}\includegraphics[scale=0.3]{MesonEx.EPSF} 
 \makebox[20mm][c]{(a)}
\end{minipage}
\begin{minipage}[t]{40mm}  
\hspace{5mm}\includegraphics[scale=0.3, bb =0 -10 400 280]{MesonGluon.EPSF}
\makebox[28mm][c]{(b)}
\end{minipage}
\begin{minipage}[t]{40mm}
 \hspace{5mm}\includegraphics[scale=0.3]{GluonCrossed.EPSF}
\makebox[30mm][c]{(c)}
\end{minipage}

The connection between the bilocal meson-diquark   action and   the ladder BSE states inspired 
an insightful alternative to the solving of the   linear homogeneous BSE.  This analysis in
Cahill, Roberts and Praschifka \cite{CRP87} involved re-formulating the ladder BSE, 
in which the boson mass of interest $M$  occurs implicitly in the quark correlators, 
as an explicit mass functional.
As an example, for the scalar  diquark state $0^+$ we find
\begin{equation}                                                                  \label{eqn;A23}   
M[\Gamma]^2=-\frac{24}{f[\Gamma]^2}\int \frac{d^4
q}{(2\pi)^4}\frac{\Gamma(q)^2}{A(q)^2q^2+B(q)^2}+\frac{9}{f[\Gamma]^2}\int d^4 x
\frac{\Gamma(x)^2}{D(x)},
\end{equation}
where $f[\Gamma]$ is a normalisation functional \cite{CRP87}. The minimisation of $M[\Gamma]$ with
respect to the boson vertex function $\Gamma$ then yields the mass $M>0$, at least as a good approximation
for the low mass mesons and diquarks. A similar expression for the pion mass functional $M_{\pi}[\Gamma]$ is
given  by (\ref{eqn;A23}), but with $9\rightarrow 9/2$ in the second term (the Landau gauge version, in the
near chiral limit, is given in (\ref{eqn;D38})). Then the minimum  of $M_{\pi}[\Gamma]=m_{\pi}=0$ arises for
$\Gamma(q)=B(q)$, as follows from the  minimisation   of $S_{CQ}[B]$  in (\ref{eqn;B18}).  However another
feature of the GCM comes into play when we use the property that all the low mass mesons and diquarks  have
approximately the same
$\Gamma(q)$, and that this is is merely the $B(q)$ function of the constituent quark correlator (an exact
result for the pion).   This intricate relationship is again a consequence of the dominance of the
constituent quark effect. Hence using
$\Gamma=B$ in (\ref{eqn;A23}) then yields an explicit value for the constituent mass of the $0^+$
diquark. Similar mass functionals for the other states are given in \cite{CRP87}.

\section{Constituent Nambu-Goldstone Mesons\label{section:NambuGoldstoneBosons}}

The properties of the pion continue to be the subject of considerable theoretical and
experimental interest in QCD studies.  The pion is an (almost) massless NG
boson and its properties are directly associated with dynamical chiral symmetry breaking and
the underlying quark-gluon dynamics.
The GCM is particularly effective in revealing the NG phenomena
that follows from the dynamical breaking of chiral symmetry \cite{RCP88,Pi94}. 
Indeed the GCM results in the complete derivation of
the Chiral Perurbation Theory (ChPT) phenomenology, but with the added feature that again the induced NG
effective action is non-local, so that the  usual non-renormalisability problems do not arise (see
sect.\ref{subsection:HiddenChiralSymmetry}). However again we must distinguish the full NG degrees of freedom
from the constituent NG modes, as again this distinction is often missing, particularly in the tDSE
formulation.
The full NG (pion) correlator ${\cal G}_{\pi}$ is the connected part of 
\begin{equation}                                                                    \label{eqn;D9} 
{\cal G}_{\pi}(x,y,z,w)=\hspace{-0.5em}\int{ D}\overline{q}{ D}q{
D}A\overline{q}(x)i\gamma_5\tau_iq(y)\overline{q}(z)i\gamma_5\tau_iq(w)
\exp(-S_{GCM}[A,\overline{q},q])
\end{equation}
or, from the hadronisation (\ref{eqn;R7}),
\begin{equation}                                                                    \label{eqn;D20}
{\cal G}_{\pi}(X,Y)=\int{ D}\pi..{ D}\overline{N}{
   D}N...\pi(X)\pi(Y)\exp(-S_{had}[\pi,...,\overline{N},N,..]) 
\end{equation}
in which $X=\frac{x+y}{2}$ and $Y=\frac{z+w}{2}$ are `centre-of-mass'  coordinates for the pion.  We note
that now the pion field appears in $S_{had}[\pi,...,\overline{N},N,..]$ in the exponent of 
(\ref{eqn;D20}),  with an effective-action    mass parameter $m_{\pi}$; see (\ref{eqn;A42}). 
It is important to clearly distinguish  this mass, together with the
equations which define its value, from the pion mass that would emerge from the evaluation of
the functional integrals in (\ref{eqn;D9}) or 
 (\ref{eqn;D20}).    Equation (\ref{eqn;D9}) or (\ref{eqn;D20})  defines the observable pion
mass, whereas the mass in the exponent  defines the constituent pion mass.   There is no
reason for these to be equal in magnitude, though they may well both be given by the
generic Gell-Mann-Oakes-Renner (GMOR)\cite{GMOR}  formula.
\begin{equation}                                                                    \label{eqn;D44}
m_{\pi}^2= \frac{12}{f_{\pi}^2}
\int\frac{d^4q}{(2\pi)^4}(m_u(s)+m_d(s))\sigma_s(q^2)+O(m^2).
\end{equation}
Recently there has been  a derivation\cite{CG95b,LaKe,CG98b} of an alternative 
 mass formula and the demonstration of its equivalence to the GMOR formula for
the constituent pion mass, which we briefly consider. That derivation exploited the  intricate interplay between the constituent pion Bethe-Salpter
equation (BSE) and the non-linear CQ equation, resulting in the new expression
\begin{equation}                                                                    \label{eqn;D6}
 m_{\pi}^2=\frac{48m}{f_{\pi}^2}  \int
\frac{d^4q}{(2\pi)^4}\left(\epsilon_s(s) \sigma_s(s)+s\epsilon_v(s)
\sigma_v(s)\right)c(s)+O(m^2),  
\end{equation} 
where $c(s)$ is the  function 
\begin{equation}                                                                    \label{eqn;D7} 
c(s)=\frac{B(s)^2}{sA(s)^2+B(s)^2},
\end{equation}
where $A(s)$ and $B(s)$ are the chiral limit of $A(s;m)$ and $B(s;m)$, and where 
 $\epsilon_s(s)$ and $\epsilon_v(s)$ are functions which specify the response of the
constituent quark correlator to the turning on of the quark current mass, defined, 
for small $m \neq 0$, by
 the  expansions in $m(s)$
\begin{equation}                                                                    \label{eqn;D34}
B(s;m)+m(s) = B(s)+ m(s).\epsilon_s(s)+O(m^2),
\end{equation}
\begin{equation}                                                                    \label{eqn;D35}
A(s,m) = A(s)+m(s).\epsilon_v(s)+O(m^2).
\end{equation}
 For large space-like $s$ we find  that $\epsilon_s
\rightarrow 1$, but for small $s$ we find that $\epsilon_s(s)$ can be significantly larger than
1. This is an infrared-region dynamical enhancement  of the quark current mass by  gluon
dressing, and indicates the strong response of the chiral limit constituent quark
correlator to the turning on of the current mass.   A plot of
$\epsilon_s(s)$ is shown in
\cite{CG95b}. 
The GCM involves the solution  of CQ integral equations (\ref{eqn;R9}) for the  constituent
correlation functions. 
Separating this CQ into its scalar and vector parts , we obtain in Landau gauge
\begin{equation}                                                                    \label{eqn;D29}
B(p^2;m)=4\int\frac{d^4q}{(2\pi)^4}D(p-q).\frac{B(q^2;m)+m(s)}
{q^2A(q^2;m)^2+(B(q^2;m)+m(s))^2},
\end{equation}
\begin{eqnarray}                                                                    \label{eqn;D30}
[A(p^2;m)-1]p^2=\frac{4}{3}\int\frac{d^4q}{(2\pi)^4}D(p-q)\left(p.q
+2\frac{q.(p-q)p.(p-q)}{(p-q)^2}\right). \nonumber  \\
\mbox{\ \ \ \ \ \ \ \ \ \  }\frac{A(q^2;m)}
{q^2A(q^2;m)^2+(B(q^2;m)+m(s))^2}.
\end{eqnarray}
which are (\ref{eqn;B15}) and (\ref{eqn;B16})  (which however are in Feynman gauge) with non-zero quark
current mass.
 We have  included, for generality,  a phenomenological momentum dependent current mass $m(s)$ for the
quarks.
The usual procedure is to introduce a cutoff $\Lambda$, and to choose an $s$
independent
$m(s)$, but with the value of that constant $m$ now $\Lambda$ dependent. This is in fact our final choice.
However our analysis supports a more general result with $m(s)$ dependent on $s$; then the GMOR
relation in (\ref{eqn;D44})  has the 
$m(s)$ included inside the integration.  Using Fourier transforms (\ref{eqn;D29}) may be written in the form,
 here for $m=0$, 
\begin{equation}                                                                    \label{eqn;D31}
D(x)=\frac{1}{4}\frac{B(x)}{\sigma_s(x)},
\end{equation}
which implies that knowledge of the chiral-limit constituent quark correlator determines the effective  gluon
correlator. Multiplying (\ref{eqn;D31}) by $B(x)/D(x)$, and using  Parseval's identity for  the RHS,
we obtain the identity 
\begin{equation}                                                                    \label{eqn;D32}
\int d^4x \frac{B(x)^2}{D(x)}=4\int\frac{d^4q}{(2\pi)^4}B(q)\sigma_s(q).
\end{equation}

The second basic equation is the ladder form BSE for the constituent pion mass-shell state,
which arises from the mesonic fluctuations about the minimum  determined by (\ref{eqn;D29}) and
(\ref{eqn;D30}),
\begin{equation}                                                                    \label{eqn;D33}
\Gamma^f(p,P)=2\int\frac{d^4q}{(2\pi)^4}D(p-q)tr_{SF}(G_+T^gG_-T^f)\Gamma^g(q,P)
\end{equation}
where $G_{\pm}=G(q\pm\frac{P}{2})$. This is (\ref{eqn;R12}) in Landau gauge for isovector NG bosons,
and only the dominant
$\Gamma=\Gamma^f T^f i\gamma_5$ amplitude is retained. The spin
trace arises from projecting onto this dominant amplitude. Here $\{T^b,b=1,..,N_F^2-1\}$ are the
generators of $SU(N_F)$, with $tr(T^fT^g)=\frac{1}{2}\delta_{fg}$.  

The BSE (\ref{eqn;D33}) is an implicit
equation for the mass shell $P^2=-m_\pi^2$.  It  has solutions {\em only} in
the time-like region $P^2 \leq 0$. Fundamentally this is ensured by (\ref{eqn;D29}) and (\ref{eqn;D30})
being the specification of an absolute minimum of an effective action after a bosonisation.  Nevertheless
the loop momentum  is kept in the space-like region $q^2 \geq 0$; this mixed metric device ensures that the
quark and gluon correlators remain close to the real space-like region where they have been most
thoroughly studied. Very little is known about these correlators in the time-like region $q^2 <
0$.  

The non-perturbative quark-gluon dynamics are expressed here in (\ref{eqn;D29}) and (\ref{eqn;D30}).
Even when  $m=0$  \hspace{2mm} (\ref{eqn;D29}) can have non-perturbative solutions with $B\neq 0$.
This is the dynamical breaking of chiral symmetry.
When $m=0$  \hspace{0.5mm} (\ref{eqn;D33}) has a solution for $P^2=0$; the Goldstone theorem
effect. For the zero linear momentum  state  $\{\vec{P}=\vec{0},P_4=0\}$  it is easily seen
that (\ref{eqn;D33}) reduces to (\ref{eqn;D29}) with $\Gamma^f(q,0)=B(q^2)$.  When $\vec{P}\neq
\vec{0} \mbox{\ \ } \mbox{\ then \ }  \Gamma^f(q,P)\neq B(q)$, and (\ref{eqn;D33})  must be solved 
for $\Gamma^f(q,P)$.

Because the pion mass $m_{\pi}$ is small when $m$ is small, we can perform an expansion of
the
$P_{\mu}$ dependence in the  kernel of (\ref{eqn;D33}). Since the analysis is Lorentz covariant we
can, without loss of validity, choose to work in the rest frame with $P=(\vec{0},im_{\pi})$,
giving $$
\Gamma(p)=\frac{1}{6}m_{\pi}^2\int\frac{d^4q}{(2\pi)^4}D(p-q)I(s)\Gamma(q)+ 
\hspace{80mm} $$
\begin{equation}                                                                    \label{eqn;D36}
\mbox{\ \ \ \ \ \ \ \ \ }
+4\int\frac{d^4q}{(2\pi)^4}D(p-q)\frac{1}{s(A(s)+m(s).\epsilon_v(s))^2+(B(s)+
m(s).\epsilon_s(s))^2}\Gamma(q)+....,
\end{equation}
where
\begin{equation}                                                                    \label{eqn;D37}
I(s)=6\left(\sigma_v^2-2(\sigma_s\sigma_s'+s\sigma_v\sigma_v')-s(\sigma_s\sigma_s''
-(\sigma_s')^2) -s^2(\sigma_v\sigma_v''-(\sigma_v')^2\right).
\end{equation}

By using Fourier transforms the  integral equation (\ref{eqn;D36}), now with explicit dependence on
$m_{\pi}$,  can be expressed in the form of a variational mass functional,
\begin{eqnarray*}
M_{\pi}[\Gamma]^2=-\frac{24}{f_{\pi}[\Gamma]^2}\int\frac{d^4q}{(2\pi)^4}\frac{\Gamma(q)^2}
{s(A(s)+m(s).\epsilon_v(s))^2+(B(s)+
m(s).\epsilon_s(s))^2}+
\end{eqnarray*}
\begin{equation}                                                                    \label{eqn;D38}
\mbox{\ \ \ \ \ \ \ \ \ \ \ \ \ \ \ \ \ \ \ \ \ \ \ \ \ \ \ \ \ \ \ \ \ \ }+
\frac{6}{f_{\pi}[\Gamma]^2}\int d^4x\frac{\Gamma(x)^2}{D(x)},
\end{equation}
which is equivalent to (\ref{eqn;A23}) for the pion (but  in the Feynman gauge),
and  in which
\begin{equation}                                                                    \label{eqn;D39}
f_{\pi}[\Gamma]^2 =
\int\frac{d^4q}{(2\pi)^4}I(s)\Gamma(q)^2.
\end{equation}
The functional derivative $\delta M_{\pi}[\Gamma]^2/\delta\Gamma(q)=0$ reproduces
(\ref{eqn;D36}).  The mass functional  (\ref{eqn;D38}) and its  minimisation is equivalent to
the constituent pion BSE in the near chiral limit.  To find an estimate for the minimum we
need only note that the change in
$m_{\pi}^2$ from its chiral limit value of zero will be of 1st order in $m$, while the change
in  $\Gamma(q)$ from its chiral limit  value $B(q^2)$ will be
of 2nd order in $m$.

Hence to obtain $m_{\pi}^2$ to lowest order in $m$, we may replace $\Gamma(q)$ by $B(q^2)$
in (\ref{eqn;D38}), and we have that the constituent pion mass is given by
\[
m_{\pi}^2=\frac{48}{f_{\pi}[B]^2}\int\frac{d^4q}{(2\pi)^4}m(s)\frac{\epsilon_s(s)B(s)+
s\epsilon_v(s)A(s)}{sA(s)^2+B(s)^2}
\frac{B(s)^2}{sA(s)^2+B(s)^2}
\mbox{\ \ \ \ \ \ \ \ \ \ \ \ \ \ \ \ \ \ \ \ \ \ \ \ \ \ \ \ \ \ \ } \]
\begin{equation}                                                                    \label{eqn;D40}
\mbox{\ \ \ \ \ \ \ \ \ \ \ \ \ \ \ \ \ \ \ \ \ \ \ \ \ \ \ \ \ \
\ }-\frac{24}{f_{\pi}[B]^2}\int\frac{d^4q}{(2\pi)^4}
\frac{B(s)^2}{sA(s)^2+B(s)^2}+\frac{6}{f_{\pi}[B]^2}\int
d^4x\frac{B(x)^2}{D(x)}+O(m^2)
\end{equation}
However the pion mass has been shown to be zero  in the chiral limit. This is confirmed
as the  two $O(m^0)$ terms in (\ref{eqn;D40})  cancel because of the identity (\ref{eqn;D32}). 
Note that it might appear that $f_{\pi}$ would contribute an extra $m$ dependence from its
kernel in (\ref{eqn;D37}).  However because the numerator in (\ref{eqn;D38}) is already of order
$m$, this extra contribution must be of higher order in $m$.  

Hence we  finally arrive at the  analytic expression, to $O(m)$, for the constituent NG boson
$(\mbox{mass})^2$ from the solution of the BSE,  
\begin{equation}                                                                    \label{eqn;D41}
m_{\pi}^2=\frac{48}{f_{\pi}[B]^2}\int\frac{d^4q}{(2\pi)^4}m(s)\frac{\epsilon_s(s)B(s)+
s\epsilon_v(s)A(s)}{sA(s)^2+B(s)^2}
\frac{B(s)^2}{sA(s)^2+B(s)^2}+O(m^2).
\end{equation}
 It would appear that expression (\ref{eqn;D41})  is manifestly different to the
conventional GMOR form in (\ref{eqn;D44}).

However here we
generalise the   identity found by Langfeld and Kettner \cite{LaKe} that
shows these forms to be equivalent. 
Inserting (\ref{eqn;D34}) and (\ref{eqn;D35}) into (\ref{eqn;D29}), and expanding in powers of
$m(s)$,  we obtain up to terms linear in
$m$, and after using  (\ref{eqn;D29}) with $m=0$ to eliminate the $O(m^0)$ terms,

\begin{eqnarray}                                       \label{eqn;D42}
m(p^2)\epsilon_s(p^2)=m(p^2)&+&4\int\frac{d^4q}{(2\pi)^4}D(p-q)
\frac{m(q^2)\epsilon_s(q^2)}{q^2A(q^2)^2+B(q^2)^2}  \nonumber \\ 
&-&4\int\frac{d^4q}{(2\pi)^4}D(p-q)\frac{B(q^2)^22m(q^2)
\epsilon_s(q^2)}{(q^2A(q^2)^2+B(q^2)^2)^2} \nonumber \\ 
&-&4\int\frac{d^4q}{(2\pi)^4}D(p-q)
\frac{B(q^2)A(q^2)2m(q^2)q^2\epsilon_v(q^2)}{(q^2A(q^2)^2+B(q^2)^2}. \nonumber \\
\end{eqnarray}

We now multiply (\ref{eqn;D42})  throughout by  $B(p^2)/(p^2A(p^2)^2+B(p^2)^2)$,
and integrate wrt $p$.  Using again the chiral limit of (\ref{eqn;D29}) there is some
cancellation of terms, and we are left with a generalised Langfeld-Kettner identity
\begin{eqnarray}                                                                    \label{eqn;D43}
2\int d^4p
\frac{B(p^2)^2}{p^2A(p^2)^2+B(p^2)^2}\left(\frac{B(p^2)m(p^2)\epsilon_s(p^2)}
{p^2A(p^2)^2+B(p^2)^2}+\frac{p^2A(p^2)m(p^2)\epsilon_v(p^2)}{p^2A(p^2)^2+B(p^2)^2}\right)=\nonumber\\
\int d^4p\frac{m(p^2)B(p^2)}{p^2A(p^2)^2+B(p^2)^2}.
\end{eqnarray}
Remarkably we see on using this identity
that  (\ref{eqn;D41}) is identical to (\ref{eqn;D44}).

\section{Constituent Nucleon\label{section:ConstituentNucleon}}

We report here progress in calculating the constituent nucleon which emerges from the GCM hadronisation as
a three constituent-quark  state, bound by the effective gluon correlator. This nucleon state is treated
as  a correlation between a constituent   quark and a constituent diquark subcorrelation in the
separable Faddeev approach. The first such computation was in 1989 \cite{BCP89} and used a rank-1
description of the scalar diquark; this 
 yielded a constituent core mass of approximately 1.3GeV, which was very close to the expected core
mass \cite{Thomas91}.  A full calculation of the nucleon core mass is particularly difficult and has yet
to be attempted, as it requires the inclusion of the constituent quark correlators and the various
constituent diquark correlators, but particularly that of the scalar $0^+$ and vector $1^+$
diquarks. This nucleon core state then has its mass further reduced by pion dressing.  As
preparation for these extensive {\it ab initio}  computations we have been monitoring the response 
of the nucleon core to the underlying low energy quark-gluon processes by computing the quark -
scalar-diquark  nucleon core state. The results are shown for GCM95 (rank 2), GCM97 (rank 3) and
GCM98 (rank 3) in Table 2.  We now briefly outline the present procedures used in these studies. 
   Working in a Euclidean metric the equation for the spin
$\frac{1}{2}^+$ nucleon form factor separable components (each a spinor) is 
\begin{eqnarray}                                                                \label{eqn;C1}
{\Psi}_i(p;P)=\frac{1}{6}\sum_{jk}\int\frac{\mbox{d}^4q}{(2\pi)^4}
\Gamma_i((p+\frac{1}{2}q+\frac{2-3\alpha}{2}P)^2)
\Gamma_j((q+\frac{1}{2}p+\frac{2-3\alpha}{2}P)^2) \nonumber\\
 .G((2\alpha-1)P-p-q)G((1-\alpha)P+q)Z_{jk}(({\alpha}P-q)^2)\Psi_k(q;P),
\end{eqnarray}
where the $\Psi_k$ are defined in terms of an arbitrary
momentum partitioning parameter $\alpha$. 
 Here the scalar diquark correlator is modelled using the  separable form
\begin{equation}                                                                \label{eqn;C2}
\Delta(q,p,P)=\sum_{ij}\Gamma_i(q)Z_{ij}(P)\Gamma_j(p),
\end{equation}
where $q$ and $p$ are the relative quark momenta, and $P$ is the diquark momentum. The matrix $Z_{ij}$
is determined from the diquark version of the ladder BS correlator (\ref{eqn;R11}) when the gluon correlator
separable expansion (\ref{eqn;R16}) is used, and only the $O(4)$ invariant part is retained. This diquark
separable expansion is not to be confused with the formal expansion in (\ref{eqn;A20}).  This  nucleon core
equation requires careful determination of its only ingredients, the quark correlator and the diquark
correlator, and particularly its vertex functions. These are determined by solving the diquark BSE using the
separable representation  of the effective gluon correlator.  No integration cutoffs are required.

We seek
solutions to  (\ref{eqn;C1})  (which being a homogeneous linear equation will only have solutions for 
particular $P^2=-M_0^2$) which give the nucleon core mass. We work 
 in the rest frame of the nucleon and accordingly set $P=({\bf 0},iM_0)$. With the above choices of
$\Gamma$, $G$, $Z$ and $P$,  (\ref{eqn;C1}) enjoys a spatial $O(3)$ symmetry.
A direct calculation shows that the integral operator commutes with
the angular momentum operator
${\bf J}={\bf L}+{\bf S}=i\frac{\partial}{\partial {\bf p}}\times {\bf p}
+\frac{1}{2}\mbox{\boldmath$\sigma$}$,
so we take the $\Psi_k$ to be one of the general $L=0$, $S=\frac{1}{2}^+$
states
\begin{equation}
\Psi_{\uparrow k}=
\left(\begin{array}{c} \left(\begin{array}{c}1\\0\end{array}\right)u_k(p) \\
  \frac{\mbox{\boldmath$\sigma$}.\bf p}{\left|{\bf p}\right|}
                       \left(\begin{array}{c}1\\0\end{array}\right)v_k(p)
\end{array}\right)\mbox{ \ \ \ \   or \ \ \ \ }
\Psi_{\downarrow k}=
\left(\begin{array}{c} \left(\begin{array}{c}0\\1\end{array}\right)u_k(p) \\
  \frac{\mbox{\boldmath$\sigma$}.\bf p}{\left|{\bf p}\right|}
                       \left(\begin{array}{c}0\\1\end{array}\right)v_k(p)
\end{array}\right),
\end{equation}
where $u_k$ and $v_k$ are functions only of $p_4$ and $\left|\bf p\right|$.
Equation (\ref{eqn;C1}) then becomes
\begin{equation}                                                        \label{eqn;C3}
\left(\begin{array}{c}u_i(p)\\v_i(p)\end{array}\right)=\sum_k{\int}\frac{d^4q}{(2\pi)^4}
  K_{ik}(p_4,\left|\bf p\right|;q_4,\left|\bf q\right|;{\bf p}.{\bf q})
  \left(\begin{array}{c}u_k(q)\\v_k(q)\end{array}\right),
\end{equation}

\begin{figure}[ht] 
\hspace{5mm}\includegraphics[scale=1.2]{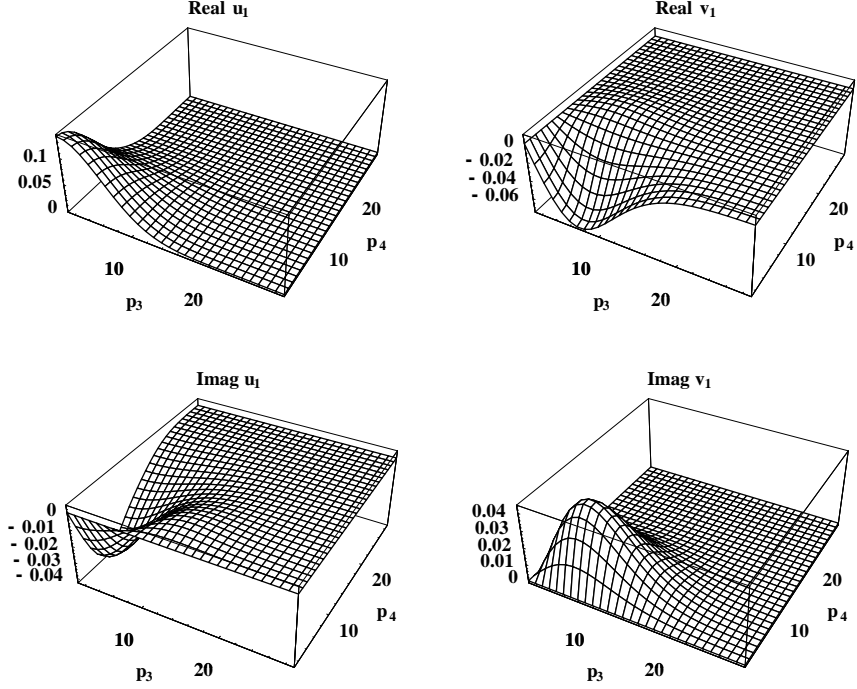}
\caption{\small{ The real and imaginary parts of the  $u_1(p_3,p_4)$ and $v_1(p_3,p_4)$ components
(corresponding to the $\Gamma_1$ in (\ref{eqn;C2}))   of the nucleon wave function  (scale: 10 units
$\equiv$ 0.57GeV.)}} 
\label{figure;Nucleonx} 
\end{figure} 

\begin{eqnarray*}
K_{ik}(p_4,|\bf p|;q_4,|\bf q|;{\bf p}.{\bf q})
 &=&\frac{1}{6}\sum_j\Gamma_i((p_4+\frac{1}{2}q_4+\frac{2-3\alpha}{2}iM_0)^2+
   |{\bf p}+\frac{1}{2}{\bf q}|^2)  \\ 
& &\mbox{\ \ \ \ \ \ \ \ \ \ \ \ \ \ \ \ \ \ \ \ \ \ }
.\Gamma_j(p{\leftrightarrow q})\tilde{G}_1\tilde{G}_2 Z_{jk}(s_3),
\end{eqnarray*}
\begin{eqnarray*}
\tilde{G}_1 &=& \frac{1}{s_1A^2(s_1)+B^2(s_1)}  \\ 
  & &\hspace{-1em}.\left(\begin{array}{cc}
   ((2\alpha\!-\!1)M_0\!+\!i(p_4\!+\!q_4))A(s_1)+B(s_1) &  
     (\left|\bf q\right|+\frac{{\bf p}.\bf q}{\left|\bf q\right|})A(s_1)\\
   (\left|\bf p\right|+\frac{{\bf p}.\bf q}{\left|\bf p\right|})A(s_1) &
     \hspace{-1em}
   (\tilde{G}_1)_{22}
                            \end{array}\right),
\end{eqnarray*}
where \mbox{\ \ \ \ } $(\tilde{G}_1)_{22}= 
[-((2\alpha\!-\!1)M_0\!+\!i(p_4\!+\!q_4))A(s_1)+B(s_1)]
          \frac{{\bf p}.\bf q}{\left|\bf q\right|\left|\bf p\right|},$
\begin{eqnarray*}
\tilde{G}_2&=&\frac{1}{s_2A^2(s_2)+B^2(s_2)} \\
             & &\hspace{-1em}.\left(\begin{array}{cc}
   ((1\!-\!\alpha)M_0\!-\!iq_4)A(s_2)+B(s_2) &   -\left|\bf q\right|A(s_2) \\
       \left|\bf q\right|A(s_2)  & -((1\!-\!\alpha)M_0\!-\!iq_4)A(s_2)+B(s_2)
                                      \end{array}\right),
\end{eqnarray*}
and the arguments of the quark and diquark correlators are
\begin{equation}                                                        
\left. \begin{array}{l}
s_1=(p_4+q_4-(2\alpha-1)iM_0)^2+\left|{\bf p}+{\bf q}\right|^2, \\
s_2=(q_4+(1-\alpha)iM_0)^2+\left|\bf q\right|^2,   \\
s_3=(q_4-i{\alpha}M_0)^2+\left|\bf q\right|^2.  \end{array}
\right\} 
\end{equation}
Equation (\ref{eqn;C3}) reduces to $n$ (the rank of the gluon correlator modelling) coupled
two-dimensional integral equations after  performing one trivial and one numerical angle
integrations.  We search for eigenvectors by introducing an eigenvalue $\lambda(M_0)$ and
changing $M_0$ until the eigenvalue $\lambda=1$. Values for $M_0$ are  in Table 2, and
$u_1(p_3,p_4)$ and $v_1(p_3,p_4)$ are shown in Fig.\ref{figure;Nucleonx}.

There have now been many quark-diquark Faddeev studies of the nucleon \cite{Buck:1992wz,
Huang:1993yd,Buck:1995ch,Hanhart:1995tc,Hellstern:1997pg,Hellstern:1998fr,HOAR98,Ebert:1996ab,
Keiner:1996at,Keiner:1996bu}, with applications to hadronic form factors \cite{Burkardt:1997gu,Ivanov:1996pz}.

\section{Pion Dressing of  Constituent Nucleon\label{section:PionDressingoftheNucleon}}

The GCM hadronisation leads directly to the formalism for dressing the constituent
nucleon by mesons as noted in sect.\ref{subsection:HadronicLaws}. The full determination of this
nucleon state is not yet completed, however we illustrate here the nature of the meson dressing
calculations. While chiral symmetry mandates the on-mass-shell  pion-nucleon coupling, 
as in (\ref{eqn;A42}),  it is clearly necessary to include the $\pi NN$ form factor
in calculating loop  processes.  The nucleon correlator ${\cal G}_N$ is then determined by 
the Euclidean metric DSE:
\begin{eqnarray}                                                                  \label{eqn;N1}
{\cal G}_N^{-1}(P)=i\backslash \!\!\!\!P +M_0+
3\frac{M M_0}{f_{\pi}^2}\int\frac{d^4q}{(2\pi)^4}\frac{1}{(P-Q)^2+m_{\pi}^2}  
\Gamma(P-Q,Q)i\gamma_5{\cal G}_N(Q)i\gamma_5\Gamma_0(Q-P,-Q), 
\end{eqnarray}
where  $\frac{M_0}{f_{\pi}}$ and $\frac{M}{f_{\pi}}$ are the core and dressed  $\pi NN$
couplings, and ${\cal G}_N(P)$ has the form
\begin{equation}                                                                  \label{eqn;N2}
{\cal G}_N(P)=(iA_N(P^2)P.\gamma+B_N(P^2)+M_0)^{-1}.
\end{equation}

\vspace{5mm}
\hspace{+5mm}\begin{minipage}[t]{50mm}

\hspace{0mm}\includegraphics[scale=0.4]{Nmass.EPSF}

\makebox[65mm][c]{(a)}
\end{minipage}
\begin{minipage}[t]{50mm}
\hspace{10mm}\includegraphics[scale=0.45]{cMITPlot.EPSF}

\makebox[85mm][c]{(b)}
\end{minipage}

\begin{figure}[ht]
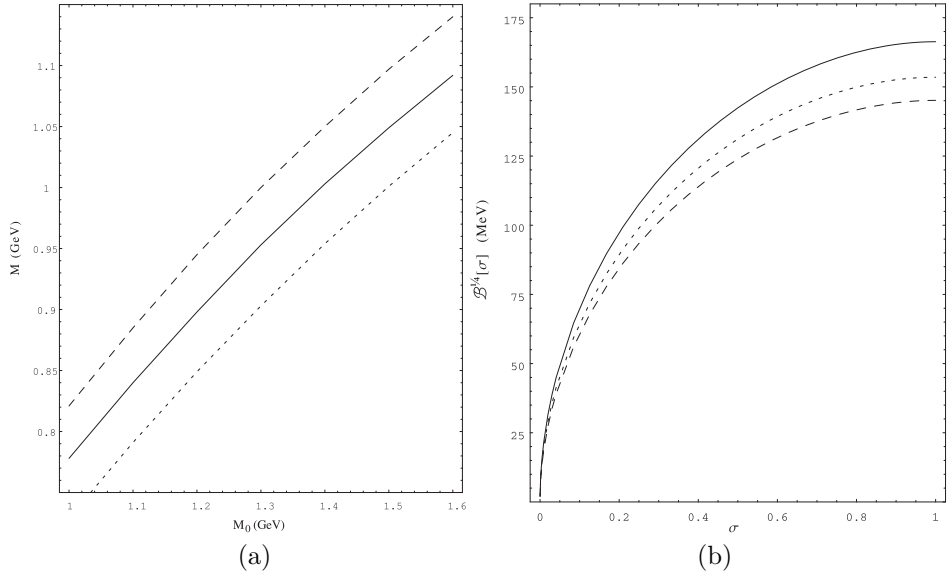
 
\vspace{-5mm}\caption{\small{ (a) Mass $M$ of the dressed nucleon for various constituent nucleon masses
$M_0$(GeV), for $\pi NN$ form factor parameter $\beta=0.55$GeV (longdash), $\beta=0.65$GeV (solid) and
$\beta=0.75$GeV (shortdash), (b) Plots of the MIT bag constant ${\cal B}^{1/4}(\sigma)$ 
for values of the scalar field $0<\sigma<1$. The curves show results
  for GCM98-solid, GCM95-dashed and
GCM97-shortdash.}
\label{figure:NucleonMass}}
\end{figure} 
Here $M_0$ is the constituent nucleon mass, while 
the mass of the dressed or physical nucleon $M$ is given by  the mass-shell condition
$P^2A_N(P^2)^2+B_N(P^2)^2|_{P^2=-M^2}=0$. This zero is situated in the time-like region and is determined
by analytic continuation from the Euclidean support $P^2 \geq 0$ in (\ref{eqn;N1}). As $M$
 occurs in one of the  couplings   when solving 
(\ref{eqn;N1}) we must find self-consistency for the value of $M$. This adds to the non-linearity
of this DSE. We model the dressed and constituent form factors  by the same separable approximation, 
\begin{equation}                                                                  \label{eqn;N3}
\Gamma(P,Q)=\Gamma_0(P,Q)=\frac{1}{(1+\frac{P^2}{\beta_1^2})(1+\frac{Q^2}{\beta_2^2})},
\end{equation}
in which $P$ is the pion momentum and $Q$ the nucleon momentum, where the parameters $\beta_1$ and
$\beta_2$ are 
computable using the nucleon structure from sect.\ref{section:ConstituentNucleon}, but here we
show results, in Fig.\ref{figure:NucleonMass}(a), for three typical values with $\beta_1=\beta_2$.  Despite
its  non-linearity
 (\ref{eqn;N1}) converges after a few iterations, indicating that the pion dressing of the
nucleon core involves only a small number of pions.  Nevertheless due to the low mass of the pion the
nucleon mass shift of typically some 300MeV  is significant, a result that is also seen in MIT
bag modelling \cite{PeAf}.

\section{Connections to Other Models\label{section:ConnectiontoOtherModels}}
By further approximations the GCM  provides a derivation of many other  models, and so makes it
possible to   link these models to QCD as illustrated in Fig.\ref{figure:Map}. A particular feature of this GCM
linking is that it can predict the values of many of the phenomenological parameters occurring in these
models, and relate their values to the underlying low-energy quark-gluon processes.   Here we briefly
indicate the connection to some of these models.

  The Nambu-Jona-Lasinio  model (NJL) is an obvious 
special case of the GCM.  Formally the NJL model \cite{Reinhardt90} is the contact
interaction limit of the GCM:
$ D_{\mu\nu}(x-y)\rightarrow g^2\delta_{\mu\nu}\delta(x-y)$  or $ D(p) \rightarrow
g^2$. But in the CQ equation  the contact limit  is undefined because it
leads to divergences in (\ref{eqn;B15}) and (\ref{eqn;B16}).  A cutoff $\Lambda$ is then always
introduced in NJL computations, which is equivalent to  using the `step-function' gluon
correlator $ D(p)
\rightarrow g^2\theta(q^2-\Lambda^2) $.  Hence the NJL model is the  GCM  but with
a box-shaped  $D(p)$, rather than a `running'  $D(p)$.

As we have already noted in sect.\ref{subsection:HiddenChiralSymmetry} the GCM provides a
comprehensive derivation of the NG sector effective action.   
This is the ChPT  effective action \cite{GL84}, but with the added insight
that all coefficients are given by explicit and convergent integrals in terms of $A$ and
$B$, which are in turn  determined by $D_{\mu\nu}$.  The higher order terms
contribute to $\pi\pi$ scattering, and the sensitivity of these to changes in $D_{\mu\nu}$ are shown in
Table 2. The GCM formalism also provides the non-local NG-baryon  effective action, and this leads
to finite values of observables, and so obviates the non-renormalisability problems that plague the
local-ChPT phenomenology.

 While the GCM hadronisation   is the  main result, at an 
intermediate stage one
obtains \cite{CR85} extended meson  Quark-Meson Coupling type models (QMC)
\cite{G88}. Applying mean field techniques to the GCM quark-meson coupling 
effective action leads to soliton type models, which have been studied in detail in
\cite{Frank} 
and the significance of the extended mesons demonstrated. From the
soliton  models a further ansatz \cite{CR85} for the form of the soliton leads to  the MIT and Cloudy
Bag Model (CBM) \cite{CBM}. In  \cite{Zuckert:1997nu}  baryons are modelled as hybrids of solitons and three
quark bound states.   An expression for the MIT Bag constant, derived from the GCM is \cite{CR85}

\begin{equation}                                                      \label{eqn;E21}
{\cal B}= \frac{12\pi^2}{(2\pi)^4}\int_0^\infty
sds\left[\mbox{ln}(\frac{A^2(s)s+B^2(s)}{A^2(s)s})
-\frac{B^2(s)}{A^2(s)s+B^2(s)}\right],
\end{equation} 
which is based on the energy density
for complete restoration of the chirally symmetric perturbative configuration inside a cavity. This bag
constant is for   core states because no meson cloud effect is included.  
With a mean field description of the pion sector  via $\sigma(x)$, 
which describes the isoscalar part of $\sigma(x)V(x)$, where $\sigma(x)$ 
is a `radial' field multiplying the NG boson field $V(x)$ (see
sect.\ref{subsection:HiddenChiralSymmetry})), ${\cal B}$ becomes, for constant $\sigma$, 

\begin{equation}                                                      \label{eqn;E22}
{\cal B}(\sigma)= \frac{12\pi^2}{(2\pi)^4}\int_0^\infty
sds\left[\mbox{ln}(\frac{A^2(s)s+\sigma^2B^2(s)}{A^2(s)s})
-\frac{\sigma^2B^2(s)}{A^2(s)s+B^2(s)}\right]
\end{equation}
which reduces to (\ref{eqn;E21}) when 
$\sigma=1$, being the non-perturbative field external to an isolated  nucleon
core.  Here $\sigma<1$ describes a partial restoration of chiral symmetry outside of
the core. Using the gluon correlator discussed in sect.\ref{section:EffectiveGluonCorrelator} we obtain
the plot of
${\cal B}^{1/4}(\sigma)$ shown in Fig.\ref{figure:NucleonMass}(b), for the three GCM gluon correlators. 
Again we emphasize that the GCM provided not only the MIT bag phenomenology but also the value of the
MIT bag constant.  Dressing of the nucleon core by mesons, using a mean field modelling, is partly
described by  a reduction in
$\sigma$ in the surface region, causing a reduction in the nucleon mass.

 However in nuclei a mean meson field   description \cite{ST94,ST95} means that
$\sigma$  is even further reduced  outside of the nucleons, and the effective bag
constant is further reduced. The $\sigma$ field can  model in part
correlated $\pi\pi$ exchanges and, along with the $\omega$ meson field, is believed
to be important in a mean field modelling of nuclei.  In \cite{JJ} it has been
argued that the  reduction of the effective bag constant  for nucleons inside
nuclei  is essential  to the recovery of features of relativistic nuclear
phenomenology.  The GCM thus allows ${\cal B}(\sigma)$ and details of relativistic
nuclear phenomenology to be directly related to the low energy quark-gluon processes
that have been extracted from low energy meson data.

\section{Conclusion                                     \label{section:Conclusion}}

The GCM  has turned out to be a very efficacious model of QCD when applied to low energy hadronic
processes. This success appears to arise from a feature of QCD that might be thought to make low energy
hadronic physics too difficult to sustain fundamental analytical models, namely the strength and number
of gluonic processes in the IR regime. However their very strength seems to lead to an IR saturation
effect in which the hadronic processes become somewhat insensitive to details of  these gluonic
processes.  This fortuitous circumstance probably also explains why there are a considerable number of
seemingly different but apparently successful hadronic models.  The GCM appears to most
successfully incorporate the manifestations of this simplifying feature of QCD.  It does so by being
itself a well-defined quantum field theory in which  the consequences of the dynamical breaking of
chiral symmetry are automatic and significant. It  also supports the powerful hadronisation analysis from
which the appropriate dynamical variables for low energy hadronic processes  naturally emerge. 
A key part of this hadronisation is a bilocal meson-diquark bosonisation of the GCM.  It is through
this non-local bosonisation that we  avoid the spurious introduction of a non-renormalisable
effective action for the hadrons.  Because of this we can uniquely relate the numerical values of 
numerous hadronic observables to the underlying modelling of the quark-gluon processes.  This procedure
is so robust that recent progress is already seeing the comparison of lattice-determined low-energy 
quark-gluon processes with those extracted from experimental data.  Until now the GCM has mainly been
applied to the meson sector, however as reported here work on an {\it ab initio} computation of the
nucleon properties within the GCM is now well advanced.  The nucleon is a complicated  state to study
not only because of its three quark character but also because the mesonic fluctuations play a
significant role. The study of the baryonic sector of the GCM will provide a rich field of phenomena
in which complex hadronic processes may be determined in the context of properly computable
quantum amplitudes devoid of the non-renormalisability problems.

\newpage


\begin{thebibliography}{10}

\bibitem{CR85}
R.~T. Cahill and C.~D. Roberts, Phys. Rev. D {\bf 32},  2419  (1985).

\bibitem{C89a}
R.~T. Cahill,  in {\em Proc. of Workshop on Diquarks}, edited by M. Anselmino
  and E. Predazzi (World Scientific, Singapore, 1989), p.\ 201.

\bibitem{C92}
R.~T. Cahill, Nucl. Phys. A {\bf 543},  63  (1992).

\bibitem{RW}
C.~D. Roberts and A.~G. Williams, Prog. Part. Nucl. Phys. {\bf 33},  477
  (1994).

\bibitem{Tand96}
P.~C. Tandy, Prog. Part. Nucl. Phys. {\bf 36},  97  (1996).

\bibitem{CG97a}
R.~T. Cahill and S.~M. Gunner, Aust. J. Phys. {\bf 50},  103  (1997).

\bibitem{Tandy}
P.~C. Tandy, Prog. Part. Nucl. Phys. {\bf 39},  117  (1997).

\bibitem{CG95a}
R.~T. Cahill and S.~M. Gunner, Phys. Lett. B {\bf 359},  281  (1995).

\bibitem{CG97b}
R.~T. Cahill and S.~M. Gunner, hep-ph/9711359.

\bibitem{CG98a}
R.~T. Cahill and S.~M. Gunner,  in {\em Proc. of Workshop on Methods of
  Non-Perturbative Field Theory}, edited by A. Schreiber, A.~W. Thomas, and
  A.~G. Williams (World Scientific, Singapore, 1999).

\bibitem{MN83}
H.~J. Munczek and A.~M. Nemirovsky.
  Phys. Rev. D {\bf 28}, 181 (1983).

\bibitem{CPB89}
R.~T. Cahill, J. Praschifka, and C.~J. Burden, Aust. J. Phys. {\bf 42},  161
  (1989).

\bibitem{C89b}
R.~T. Cahill, Aust. J. Phys {\bf 42},  171  (1989).

\bibitem{Reinhardt90}
H. Reinhardt, Phys. Lett. B {\bf 244},  316  (1990).

\bibitem{Ebert:1991qg}
D. Ebert, Hadronisation of QCD, Invited talk given at Rencontres de Moriond
  Mtg. High Energy Hadronic Interactions, Les Arcs, France, Mar 17-23, 1991.

\bibitem{Ebert:1997hr}
D. Ebert and T. Jurke, hep-ph/9710390.

\bibitem{Ebert:1992zq}
D. Ebert and L. Kaschluhn, Phys. Lett. {\bf B297},  367  (1992).

\bibitem{Ebert:1993xd}
D. Ebert, A.~A. Belkov, A.~V. Lanyov, and A. Schaale, Int. J. Mod. Phys. {\bf
  A8},  1313  (1993).

\bibitem{Ebert:1997ey}
D. Ebert, hep-ph/9710507.

\bibitem{Lich}
D.~B. Lichtenberg, in {\em Proc. of Workshop on Diquarks} (World Scientific,
  Singapore, 1989).

\bibitem{Ida}
M. Ida and R. Kobayashi, Prog. Theor. Phys. {\bf 36},  846  (1966).

\bibitem{BRvS}
A. Bender, C.~D. Roberts, and L. von Smeckal, Phys. Lett. B {\bf 380},  7
  (1996).

\bibitem{PRC87a}
J. Praschifka, C.~D. Roberts, and R.~T. Cahill, Int. J. Mod. Phys. A {\bf 2},
  1797  (1987).

\bibitem{PRC87b}
J. Praschifka, C.~D. Roberts, and R.~T. Cahill, Phys. Rev. D {\bf 36},  209
  (1987).

\bibitem{RCP88}
C.~D. Roberts, R.~T. Cahill, and J. Praschifka, Ann. Phys. {\bf 188},  20
  (1988).

\bibitem{Pi94}
C.~D. Roberts, R.~T. Cahill, M.~E. Sevior, and N. Iannella, Phys. Rev. D {\bf
  49},  125  (1994).

\bibitem{CRP89}
R.~T. Cahill, C.~D. Roberts, and J. Praschifka, Aust. J. Phys. {\bf 42},  129
  (1989).

\bibitem{RCP89}
C.~D. Roberts, R.~T. Cahill, and J. Praschifka, Int. J. Mod. Phys. A {\bf 4},
  719  (1989).

\bibitem{Mitchell:1994jj}
K.~L. Mitchell, P.~C. Tandy, C.~D. Roberts, and R.~T. Cahill, Phys. Lett. {\bf
  B335},  282  (1994).

\bibitem{Mitchell:1996dn}
K.~L. Mitchell and P.~C. Tandy, Phys. Rev. {\bf C55},  1477  (1997).

\bibitem{Frank}
M.~R. Frank and P.~C. Tandy, Phys. Rev. C {\bf 46},  338  (1992).

\bibitem{Frank:1993ye}
M.~R. Frank and P.~C. Tandy, Phys. Rev. {\bf C49},  478  (1994).

\bibitem{Hecht:1997uj}
M.~B. Hecht and B.~H.~J. McKellar, Phys. Rev. {\bf C57},  2638  (1998).

\bibitem{Alkofer:1995gu}
R. Alkofer, A. Bender, and C.~D. Roberts, Int. J. Mod. Phys. {\bf A10},  3319
  (1995).

\bibitem{Frank:1996yb}
M.~R. Frank and T. Meissner, Phys. Rev. {\bf C53},  2410  (1996).

\bibitem{KK97} 
D. Klab{\"u}car and D. Kekez, hep-ph/9710206.

\bibitem{MR}
P. Maris and C.~D. Roberts, Phys. Rev. D {\bf 56},  3369  (1997).

\bibitem{JM}
P. Jain and H.~J. Munczek, Phys. Rev. D {\bf 48},  5403  (1993).

\bibitem{KK}
D. Kekez and D. Klab{\"u}car, Phys. Lett. B {\bf 387},  14  (1996).

\bibitem{LSWP}
D.~B. Leinweber, J.~I. Skullerud, A.~G. Williams, and C. Parrinello,
  hep-lat/9803015.

\bibitem{Marenzoni} P. Marenzoni,  G.  Martinelli, and  N. Stella, {Nucl.Phys.} B {\bf 455},
339(1995). 

\bibitem{Skul98}
J.~I. Skullerud, Nucl. Phys. Proc. Suppl. {\bf 63},  242  (1998).

\bibitem{PCR88}
J. Praschifka, R.~T. Cahill, and C.~D. Roberts, Int. J. Mod. Phys. A {\bf 3},
  1595  (1988).

\bibitem{Hellstern:1997nv}
G. Hellstern, R. Alkofer, and H. Reinhardt, Nucl. Phys. {\bf A625},  697
  (1997).

\bibitem{CRP87}
R.~T. Cahill, C.~D. Roberts, and J. Praschifka, Phys. Rev. D {\bf 36},  2804
  (1987).

\bibitem{GMOR}
M. Gell-Mann, R. Oakes, and B. Renner, Phys. Rev. {\bf 175},  2195  (1968).

\bibitem{CG95b}
R.~T. Cahill and S.~M. Gunner, Mod. Phys. Lett A {\bf 39},  3051  (1995).

\bibitem{LaKe}
K. Langfeld and C. Kettner, Mod. Phys. Lett. A {\bf 41},  1331  (1996).

\bibitem{CG98b}
R.~T. Cahill and S.~M. Gunner, Aust. J. Phys. {\bf 51},  509  (1998).

\bibitem{BCP89}
C.~J. Burden, R.~T. Cahill, and J. Praschifka, Aust. J. Phys. {\bf 42},  147
  (1989).

\bibitem{Thomas91}
A.~W. Thomas, Aust. J. Phys. {\bf 44},  173  (1991).

\bibitem{Buck:1992wz}
A. Buck, R. Alkofer, and H. Reinhardt, Phys. Lett. {\bf B286},  29  (1992).

\bibitem{Huang:1993yd}
S-zhou Huang and J. Tjon, Phys. Rev. {\bf C49},  1702  (1994).

\bibitem{Buck:1995ch}
A. Buck and H. Reinhardt, Phys. Lett. {\bf B356},  168  (1995).

\bibitem{Hanhart:1995tc}
C. Hanhart and S. Krewald, Phys. Lett. {\bf B344},  55  (1995).

\bibitem{Hellstern:1997pg}
G. Hellstern, R. Alkofer, M. Oettel, and H. Reinhardt, Nucl. Phys. {\bf A627},
  679  (1997).

\bibitem{Hellstern:1998fr}
G. Hellstern, M. Oettel, R. Alkofer, and H. Reinhardt, hep-ph/9805054.

\bibitem{HOAR98}
G. Hellstern, M. Oettel, R. Alkofer, and H. Reinhardt, hep-ph/9805393.

\bibitem{Ebert:1996ab}
D. Ebert, T. Feldmann, C. Kettner, and H. Reinhardt, Int. J. Mod. Phys. {\bf
  A13},  1091  (1998).

\bibitem{Keiner:1996at}
V. Keiner, Phys. Rev. {\bf C54},  3232  (1996).

\bibitem{Keiner:1996bu}
V. Keiner, Z. Phys. {\bf A354},  87  (1996).

\bibitem{Burkardt:1997gu}
M. Burkardt, M.~R. Frank, and K.~L. Mitchell, Phys. Rev. Lett. {\bf 78},  3059
  (1997).

\bibitem{Ivanov:1996pz}
M.~A. Ivanov, M.~P. Locher, and V.~E. Lyubovitskii, Few Body Syst. {\bf 21},
  131  (1996).

\bibitem{PeAf}
B.~C. Pearce and I.~R. Afnan, Phys. Rev. C {\bf 34},  191  (1986).

\bibitem{GL84}
J. Gasser and H. Leutwyler, Ann. Phys. {\bf 158},  142  (1984).

\bibitem{G88}
P.~A.~M. Guichon, Phys. Lett. B {\bf 200},  235  (1988).

\bibitem{CBM}  
 A.~W. Thomas, S. Th\'{e}berge and  G.~A. Miller,  
 Phys. Rev. D{\bf 24}, 216 (1981). 

\bibitem{Zuckert:1997nu}
U. Zuckert, R. Alkofer, H. Weigel, and H. Reinhardt, Phys. Rev. {\bf C55},
  2030  (1997).

\bibitem{ST94}
K. Saito and A.~W. Thomas, Phys. Lett. B {\bf 327},  9  (1994).

\bibitem{ST95}
K. Saito and A.~W. Thomas, Phys. Rev. C {\bf 52},  2789  (1995).

\bibitem{JJ}
X. Jin and B.~K. Jennings, Phys. Lett. B {\bf 374}, 13 (1996).

\end{thebibliography}
 


\end{document}